\documentstyle[graphics,epsfig,wrapfig]{cjpsuppl}
\newcommand{\la}{\langle}
\newcommand{\ra}{\rangle}
\newcommand{\di}{ {\rm d} }
\begin{document}

 \title{
 Transversity distribution in spin asymmetries in
 semi-inclusive DIS and in the Drell-Yan process}
\authori{\underline{A.~V.~Efremov}\footnote{
Partially supported by grants
RFBR 03-02-16816 and DFG-RFBR 03-02-04022.}
}
\addressi{Joint Institute for Nuclear Research, Dubna 141980,
Russia}
\authorii{K.~Goeke and P.~Schweitzer}
\addressii{Institut f\"ur Theoretische Physik II,
Ruhr-Universit\"at Bochum, Germany}
 \authoriii{}   \addressiii{}
 \authoriv{}    \addressiv{}
 \authorv{}     \addressv{}
 \authorvi{}    \addressvi{}
 \headtitle{Transversity in SIDIS and Drell-Yan process}
 \headauthor{A.~V.~Efremov, K.~Goeke and P.~Schweitzer}
 \lastevenhead{author: A.~V.~Efremov\ldots}
 \refnum{}
 \daterec{} 
 \suppl{A}  \year{2005} \setcounter{page}{1}
 \maketitle

\begin{abstract}
A review is given on single spin asymmetries in deep inelastic
semi-inclusive scattering (SIDIS) and their possible theoretical
understanding in the framework of QCD-induced factorization
approach, wherefore predictions for transversity from the chiral
quark soliton model are used. Arising difficulties to interpret
most recent SIDIS data, and the possibility to access the
transversity distribution in the Drell-Yan process are discussed.
\end{abstract}

\section{Introduction}
\label{sect1}

It is well known that three most important (twist-2) elements of
the parton density matrix in a nucleon are the non-polarized
parton distributions functions (PDF) $f_1(x)$, the longitudinal
spin distribution  $g_1(x)$ and the transverse spin distribution
(transversity) $h_1(x)$ \cite{transversity}. The first two have
been successfully measured experimentally in classical deep
inelastic scattering (DIS) experiments but the measurement of the
last one is especially difficult since it belongs to the class of
the so-called chiral-odd structure functions and can not be seen
there.

The non-polarized PDF's have been measured for decades and are
well known in wide range of $x$ and $Q^2$. Their behaviour in
$Q^2$ is well described by the QCD evolution equation and serves
as one of the main sources of $\alpha_s(Q^2)$ determination.

The longitudinal spin PDF's drew common attention during last
decade in connection with the famous "Spin Crisis", i.e.
astonishingly small portion of the proton spin carried by quarks
(see \cite{ael} and references therein).  The most popular
explanation of this phenomenon is large contribution of the gluon
spin $\Delta G(x)$.  The direct check of this hypothesis is one
of the main problems of running dedicated experiments like
COMPASS at CERN and RHIC at BNL. Even now, however, there is some
indication to a considerable value of $\Delta G(x)$ coming from
the $Q^2$ evolution of the polarized PDF's \cite{leader99} and
from the first direct experimental probe of $\Delta G(x)$ by
HERMES collaboration \cite{hermes00} with the result $\Delta
G(x)/G(x)=0.41\pm0.18$ in the region $0.07<x<0.28$. The latter is
in reasonable agreement with large $N_c$ limit prediction
\cite{efremov00} $\Delta G(x)/G(x)\approx 1/N_c$ for not very
small $x$.

Another problem here is the sea quark spin asymmetry. It is
usually assumed in fitting the experimental data that $\Delta\bar
u=\Delta\bar d =\Delta\bar s$. This {\it ad hoc} assumption
however contradicts with large $N_c$ limit prediction $\Delta\bar
u\approx-\Delta\bar d$ \cite{efremov00}. This was previously
discovered in the instanton model \cite{DK} and supported by
calculations in the chiral quark soliton model ($\chi$QSM)
\cite{Diakonov:1996sr,Dressler00}. An indication to a nonzero
value for $\Delta \bar{u} -\Delta \bar{d}$  was also observed in
\cite{MY}.

Concerning the transversity distribution it was completely
unknown experimentally till recent time. The only information
comes from the Soffer inequality \cite{Soffer:1995ww}
$|h_1(x)|\le{1\over2}[f_1(x)+ g_1(x)]$ which follows from density
matrix positivity. To access these chiral-odd structure functions
one needs either to scatter two polarized protons and to measure
the transversal spin correlation $A_{NN}$ in Drell-Yan process
that is the problem for running RHIC and future PAX (GSI)
experiments (see e.g. \cite{Efremov:2004qs,Anselmino:2004ki})
or to know the transverse polarization of a quark scattered from
transversely polarized target. There are several ways to do this:
\begin{enumerate}
\item
To measure the polarization of a self-analyzing hadron to which
the quark fragments in a semi inclusive DIS (SIDIS), e.g.
$\Lambda$-hyperon.  The drawback of this method however is a
rather low rate of quark fragmentation into $\Lambda$-particle
($\approx 2\%$) and especially that it is mostly sensitive to
$s$-quark polarization. Also the polarization transfer from
parton to $\Lambda$-hyperon is unknown.
\item
To measure a transverse handedness in multi-particle parton
fragmentation \cite{hand}, i.e. the correlation of the parent quark spin
4-vector $s_\mu$ and jet particle momenta $k_i^\nu$,
$\epsilon_{\mu\nu\sigma\rho}s^\mu k_1^\nu k_2^\sigma k^\rho$
($k=k_1+k_2+k_3+\cdots$ is a jet 4-momentum).
\item
To use a new spin dependent T-odd parton fragmentation function
(PFF) \cite{Mulders:1995dh,Boer:1997nt,muldz} responsible for the
left-right asymmetry in one particle fragmentation of
transversely polarized quark relative to the quark momentum--spin
plane. (The so-called "Collins asymmetry" \cite{collins}.)
\end{enumerate}

The last two methods are comparatively new and only in the last
years some experimental indications to the transversal handedness
\cite{czjp99} and to the T-odd PFF \cite{todd} have appeared.

Concerning the new PDF's and PFF's. Analogous of PDF's $f_1,\
g_1$ and $h_1$ are the PFF's $D_1,\ G_1$ and $H_1$, which
describe the fragmentation of a non-polarized quark into a
non-polarized hadron  and a longitudinally or transversely
polarized quark into a longitudinally or transversely polarized
hadron, respectively. These PFF's are integrated over the
transverse momentum ${\bm p}_{h\perp}$ of a hadron with respect
to a quark. With ${\bm p}_{h\perp}$ taken into account, new PFF's
arise. Using the Lorentz- and P-invariance one can write in the
leading twist approximation 8 independent spin structures. Most
spectacularly it is seen in the helicity basis where one can
build 8 twist-2 combinations, linear in spin matrices of the
quark and hadron {$\bm\sigma$}, ${\bm S}$ with momenta ${\bm
k'}$, ${\bm p_h}$.

These PFF can be used to extract the information on the proton
transversity distribution from azimuthal asymmetries in SIDIS
with hadron production (pions and kaons) on a polarized nucleon
target
\be
\label{reaction}
l+\vec N \to l'+ h + X
\ee
recently observed by HERMES
\cite{Airapetian:1999tv}--\cite{Avetisyan:2004uz} and CLAS
\cite{Avakian:2002qp,Avakian:2003pk} collaborations.

In this review we present the results of the works
\cite{Efremov:2000za}--\cite{Efremov:2002ut}, \cite{Efremov:2004qs}
on spin asymmetries in SIDIS and the Drell-Yan process, which are
based on predictions for the transversity distributions from the
$\chi$QSM \cite{Schweitzer:2001sr}, c.f.\  also the overview
given in Ref.~\cite{Efremov:2004ph}.


\section{SIDIS kinematics and azimuthal asymmetries}
\label{sect-kin-asym}

In the framework of the parton model the squared matrix element
modulus of the  process (\ref{reaction}) is represented by
\begin{figure}[htb]
\begin{minipage}[t]{.40\textwidth}
\includegraphics[width=0.85\textwidth]{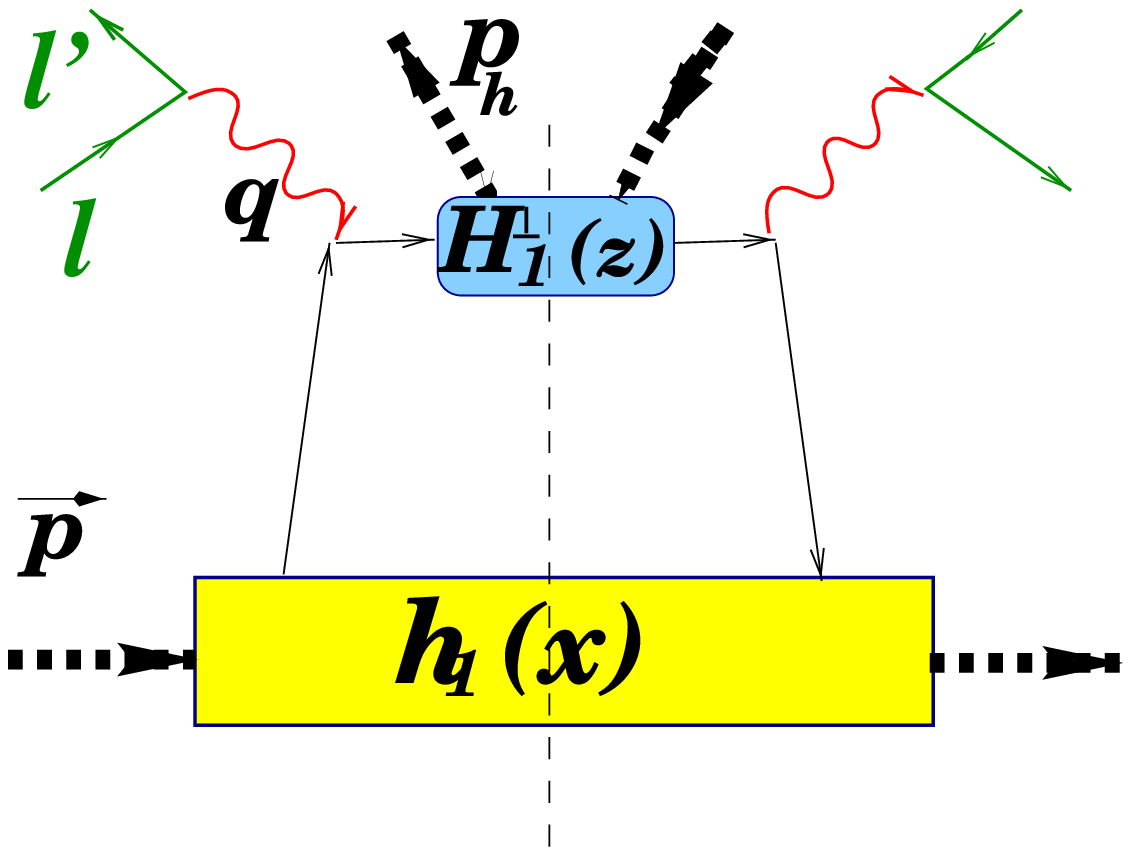}\\[-10mm]
\caption{The squared modulus of the matrix element of the
process (\ref{reaction}) in the parton model summed over states
$X$. The $H_1^\perp(z)$ and $h_1(x)$ are  examples of PFF and
PDF, respectively. \hspace{4.1cm} {}}
\label{diagram}
\end{minipage}
\hfil
\begin{minipage}[t]{.55\textwidth}
\raisebox{-7ex}{\includegraphics[width=
\textwidth,height=45mm]{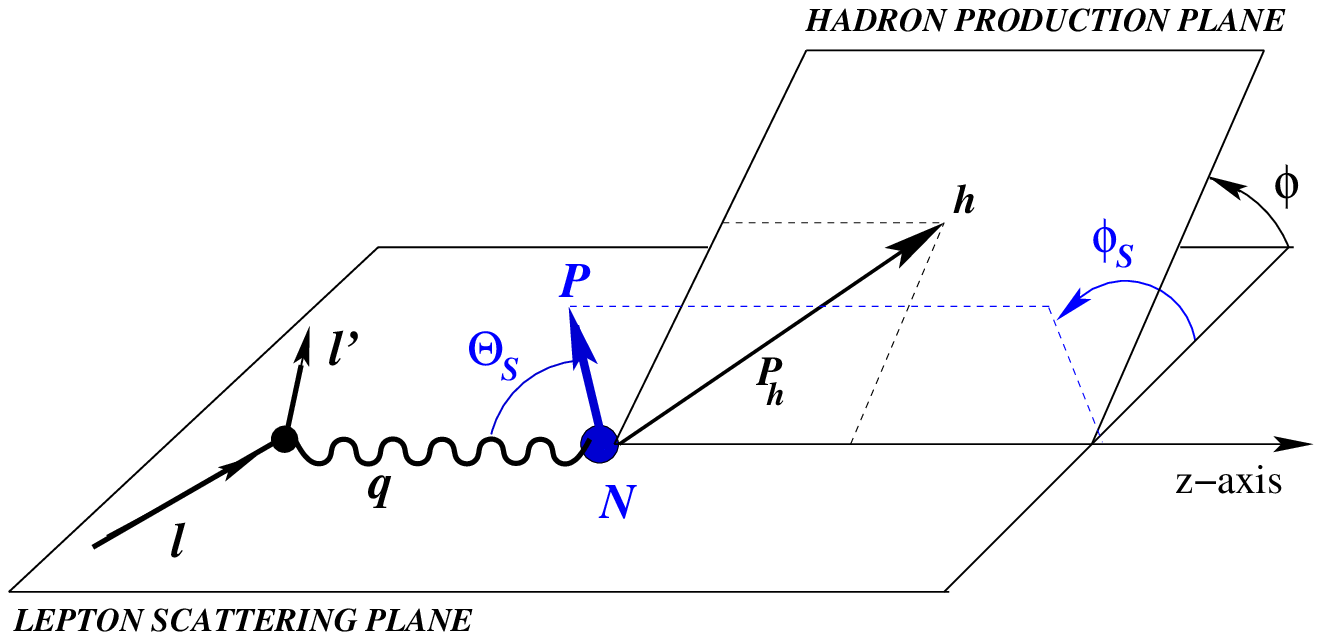}} \caption{Kinematics of
the process (\ref{reaction}).} \label{kinsidis}
\end{minipage}
\end{figure}
the diagram in Fig.\ref{diagram}  and can be written as a sum of
products of $x$-dependent quark distribution functions in a
nucleon, $x=\frac{Q^2}{2p\cdot q}$, with $q=l-l'$, $Q^2=-q^2$,
and $z$-dependent quark fragmentation functions of scattered
quark into hadron $h$, $z=\frac{p\cdot p_h}{p\cdot q}$. The
kinematics of the process (\ref{reaction}) is presented in
Fig.\ref{kinsidis}.

The cross section of the semi-inclusive production  of hadrons by
polarized leptons on the polarized target is a linear function of
the longitudinal lepton beam polarization, $P_l$, and the target
polarization, $P$, with  longitudinal,  $P_L$, and transversal,
$P_T$, components relative to a virtual photon momentum $\vec q$
in laboratory r.f.:

\be
\label{xsect}
d\sigma=d\sigma_{00}+P_l d\sigma_{L0}+
P_L(d\sigma_{0L}+P_l d\sigma_{LL})
+|P_T|(d\sigma_{0T}+P_l d\sigma_{LT}).
\ee

For the target polarization $P_\|$, longitudinal relative to the
lepton beam, the transverse component is equal to
$|P_T|=P_\|\sin\theta_\gamma$, where $\theta_\gamma$ is the angle
of the virtual photon momentum $\vec q$ relative to the lepton
beam,
\be
\label{PT}
\sin\theta_\gamma\approx
2\frac{M}{Q}x\sqrt{1-y}\,,
\ee
$y=\frac{p\cdot q}{p\cdot l}$ and $M$ is the nucleon mass.

In the parton model each of the partial cross sections
contributing to the Eq. (\ref{xsect}) is characterized by the
specific dependence on the azimuthal angle of an outgoing hadron,
$\phi$, and on the azimuthal angle of transversal component of
the target polarization vector\footnote{For longitudinal target
polarization $P_\|$ the angle $\phi_S=0$ or $\pm\pi$}, $\phi_S$,
relative to the lepton scattering plane (see Fig.\ref{kinsidis})
times the definite product of PDF and PFF summed over quark and
antiquark flavour times its charge squared. Namely, contributions
to (\ref{xsect}) for each quark and antiquark flavour up to the
order ${\cal O}(M/Q)$ 
have the forms\footnote{We use the
notations of the work \cite{Mulders:1995dh,Boer:1997nt,muldz}.
The letters $g_1(G_1),\ h_1(H_1),\ f_1(D_1)$ indicate twist-2 PDF
(PFF) with longitudinally, transversally polarized or unpolarized
partons, subscripts $L,\ T$ indicate the polarization of hadron
and superscripts $\perp$ indicates a $p_\perp$-dependence. Note
that very recently further structures have been introduced and
discussed \cite{Afanasev:2003ze} that could contribute the
longitudinally polarized cross sections. }
\cite{Mulders:1995dh,Boer:1997nt}:

\vspace{12mm}
\be
\label{pdfpff}
~
\ee

\vspace{-20mm}
\begin{tabular}{ll}
$d\sigma_{00}\propto$& $ xf_1(x)D_1(z)+
      xh_1^\perp(x) H_1^\perp(z)\cos2\phi -
      \frac{M}{Q}x^2f^\perp(x) D_1(z)\cos\phi$\,, \\
$d\sigma_{L0}\propto$& $\frac{M}{Q}x^2e(x)H_1^\perp(z)\sin\phi
+\frac{M}{Q}x^2h_1^\perp(x)E(z)\sin\phi$\,, \\
$d\sigma_{0L}\propto$&
{$xh_{1L}^\perp(x)
    H_1^\perp(z)\sin 2\phi $}
    {$+\frac{M}{Q}x^2h_{L}(x)
    H_1^\perp(z)\sin\phi$}\,,\\
$d\sigma_{LL}\propto$& $xg_1(x)D_1(z)+
     \frac{M}{Q}x^2g^\perp_L(x)D_1(z)\cos\phi$\,, \\
$d\sigma_{0T}\propto$&
{$xh_1(x)H_1^\perp(z)\sin(\phi+\phi_S)$}
       $+xh_{1T}^\perp(x) H_1^\perp(z)\sin(3\phi-\phi_S)$\\
    &{$+xf_{1T}^\perp(x)
      D_1(z)\sin(\phi-\phi_S)$}\,,\\
$d\sigma_{LT}\propto$& $xg_{1T}(x)D_1(z)\cos(\phi-\phi_S)$\,,\\[2mm]
\end{tabular}

\noindent
where
\begin{itemize}

\item[]
$f_1(x)\equiv q(x)$ is PDF of non-polarized quarks in a
non-polarized target,

\item[]
$g_1(x)\equiv\Delta q(x)$ is PDF of the longitudinally polarized
quarks in the longitudinally polarized target,
\item[]$g_{1T}(x)$
is the same as $g_1(x)$ but in the transversally polarized
target,

\item[]
$h_1(x)$ is PDF of the transversally polarized quark with
polarization  parallel to that one of a transversally polarized
target (so-called transversity),

\item[]
$h^{\perp}_{1L,T}(x)$ is PDF of the transversally polarized quark
with polarization perpendicular to the hadron polarization in the
longitudinally or transversally polarized target,

\item[]$h^{\perp}_1(x)$
is PDF of the transversally polarized quark in the non-polarized
target,

\item[]$f^{\perp}_{1T}(x)$
is PDF responsible for a left-right asymmetry in the distribution
of the non-polarized quarks in the transversally polarized target
(so-called  Sivers PDF \cite{Sivers:1989cc}),

\item[]$D_1(z)$
is PFF of the non-polarized quark in the non-polarized or
spinless produced hadron,

\item[]$H^{\perp}_1(z)$
is PFF responsible for a left-right asymmetry in the
fragmentation of a transversally polarized quark into a
non-polarized or spinless produced hadron (so-called Collins
PFF \cite{collins}).
\end{itemize}

The others ($E,\ e,\ g_L^\perp,\ h_L,\ f^\perp$) are the twist-3
functions entering with a factor $M/Q$. They have no definite
probabilistic interpretation, but are connected (except for
$e(x)$ and $E(z)$) to the above listed functions by the
approximate integral relations of the Wandzura-Wilczek type. For
example \cite{Mulders:1995dh,Boer:1997nt}

\be
\int d^2k_\perp\left(\frac{k_\perp^2}{2M^2}\right)h_{1L}^{\perp}(x,k_\perp)
\equiv h_{1L}^{\perp(1)}(x)=
-(x/2)h_L(x)=-x^2\int\limits_x^1d\xi h_1(\xi)/\xi^2.
\label{wwform}
\ee
Besides, in formula (\ref{pdfpff}) each term should be multiplied
by known kinematic factors depending on $y$, $\la k_\perp\ra$ and
$\la p_{h\perp}\ra$ (assuming Gaussian distribution) which are
omitted for simplicity.

The azimuthal asymmetries are defined as
\be\label{asim}
A_{BA}^{W(\phi,\phi_S)}(x,z,h) =
\frac{\displaystyle
\int\!\!\di y\,\di\phi\,\di\phi_S\,W(\phi,\phi_S)\,\left(
\frac{1}{P_+}\,\frac{\di^4\sigma_D^+}{\di x\,\di y\,\di z \di\phi}-
\frac{1}{P_-}\,\frac{\di^4\sigma_D^+}{\di x\,\di y\,\di z \di\phi}\right)}
{\;\;\;\;\;\;\;\displaystyle
\frac{1}{2}\int\!\!\di y\,\di\phi\,\left(
\frac{\di^4\sigma_D^+}{\di x\,\di y\,\di z\di\phi}+
\frac{\di^4\sigma_D^-}{\di x\,\di y\,\di z\di\phi}\right)}\;\;,
\ee
where $W(\phi,\phi_S)$ is an angular dependent weight from Eqs.
(\ref{pdfpff}) and $P_\pm$ denotes the target polarization
modulus. The subscripts B and A are 0, L or T for the
unpolarized, longitudinally or transversally polarized beam or
target (relative to the virtual photon direction).

It is clear that for the transversal target polarization one can
separate all the components contained in (\ref{pdfpff}) carrying
out the Fourier-analysis with respect to the angles $\phi$ and
$\phi_S$ plus change of the polarization sign (using
anti-symmetrization and symmetrization). For example, for
separation of the term with the pure transversity contribution
$h_1(x)H_1^\perp(z)$ it is enough to calculate the average value
$\la\sin(\phi+\phi_S)\ra$ and for separation of the  Sivers
function $f_{1T}^{\perp}(x)D_1(z)$ it is enough to find the
average value $\la\sin(\phi-\phi_S)\ra$.

For the longitudinal target polarization the total separation is
impossible since $\phi_S=\pm\pi$ or $0$ (see Fig.\ref{kinsidis})
and several different mechanisms can produce the
$\sin\phi$-asymmetry.  However it is possible to single out the
function $h_{1L}^\perp(x)$,  which is connected with transversity
via Eq. (\ref{wwform}), by measuring $\la\sin(2\phi)\ra$.

\section{Collins PFF}
\label{sect-Collins}

As it is seen from (\ref{pdfpff}) the Collins PFF that describes a
left--right asymmetry in the fragmentation of a transversely
polarized quark is especially interesting since it enters any
term connected with the transversity. The corresponding term of
fragmentation has the structure
$$
H_1^\perp\mbox{\bm\sigma}({\bm k'}\times
{\bm p}_{h\perp})/k'\la p_{h\perp}\ra,
$$
where $H_1^\perp$ is a function of the longitudinal momentum
fraction $z$ and hadron transverse momentum  $p_{h\perp}$. The
$\la p_{h\perp}\ra$ is the averaged transverse momentum of the
final hadron\footnote{Notice different normalization factor
compared with \cite{muldz}, $\la p_{h\perp}\ra$ instead of
$M_h$.}. Since the $H_1^\perp$ term is chiral-odd, it makes
possible to measure the proton transversity distribution $h_1$ in
semi-inclusive DIS from a transversely polarized target by
measuring the left-right asymmetry of forward produced pions. The
ratio $H_1^{\perp}\over D_1$ serves as analyzing power of
the Collins effect.

The problem is that, first, this function was completely unknown
till recent time both theoretically and experimentally. Second,
the function $H_1^\perp$ is the so-called T-odd fragmentation
function: under the naive time reversal ${\bm p_h},\ {\bm k'},\
{\bm S}$ and $\bm\sigma$ change sign, which demands a purely
imaginary (or zero) $H_1^\perp$ in the contradiction with naive
hermiticity. This, however, does not mean the break of
T-invariance but rather the presence of an interference of
different channels in forming the final state with different
phase shifts, like in the case of single spin asymmetry
phenomena\footnote{In this aspect they are very different from
the T-odd PDF's like $f_{1T}^\perp$ or $h_1^\perp$ which can not
exist since they are purely real. Interaction among initial
hadrons which could bring an imaginary part breaks the
factorization and the whole parton picture. Recently however it
was stated \cite{BMT} that effectively the necessary imaginary
phase shift can appear due to propagation of the scattered parton
in gluon field of the nucleon remnant. Since this phase shift
depends on the subprocess the corresponding PDF is, generally
speaking, not universal. In particular, it was shown that the
Sivers PDF in one loop approximation, where its factorization
seems proven \cite{Ji:2004wu,Collins:2004nx}, should have
opposite sign in Drell-Yan and SIDIS processes. Also 
these functions are suppressed in $\chi$QSM (see footnote
\ref{chiral-t-odd}).}

\cite{gasior}. A calculations of this function in simple
perturbative chiral Manohar-Georgi model can be found in
\cite{Bacchetta:2002es}.

\begin{wrapfigure}[16]{R}{6.2cm}
\vspace{-10mm}
\includegraphics[width=6.2cm,height=6.0cm]{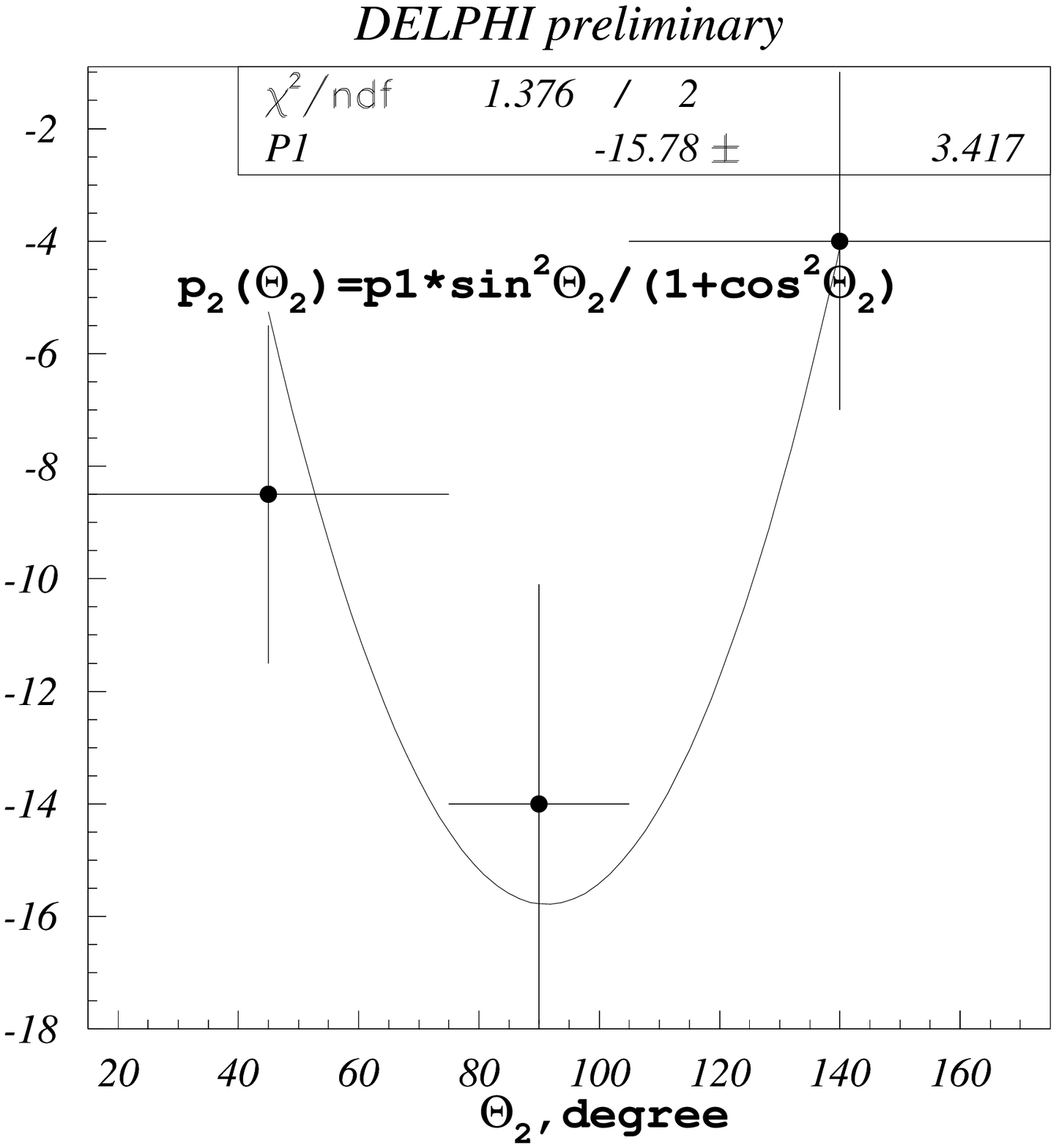}\\[-12mm]
\caption{ The $\theta_2$-dependence of the value \newline
${}\;\;\;\;\;\;\;
P_2={6\over\pi}\left|{\la H_1^{\perp}\ra\over\la D_1\ra}\right|^2
\overline{C}_{TT}{\sin^2\theta_2\over 1+\cos^2\theta_2}$ (in ppm).}
\label{fig4}
\end{wrapfigure}

Meanwhile, the data collected by DELPHI (and other LEP
experiments) give a possibility to measure $H_1^\perp$.  The
point is that despite the fact that the transverse polarization
of quarks in $e^+e^-\to Z^0\to q\bar q$ is very small
($O(m_q/M_Z)$), there is a non-trivial correlation $C^{q\bar
q}_{TT}$ between transverse spins of a quark and an antiquark. In
the Standard Model: $C^{q\bar q}_{TT}=
{(v_q^2-a_q^2)/(v_q^2+a_q^2)}$, which are at $Z^0$ peak:
$C_{TT}^{u,c}\approx -0.74$ and $C_{TT}^{d,s,b}\approx -0.35$.
With the production cross section ratio $\sigma_u/\sigma_d=0.78$
this gives for the average over flavours value $\la
C_{TT}\ra\approx -0.5$.

The transverse spin correlation results in a peculiar azimuthal
angle dependence of produced hadrons, if the T-odd fragmentation
function $H_1^\perp$ does exist~\cite{collins,colpsu}.  A simpler
method has been proposed by the Amsterdam group \cite{muldz}.
They predict a specific azimuthal behaviour of a hadron in a jet
about the axis in direction of another hadron in the opposite
jet
\footnote{The factorized Gaussian form of $p_{h\perp}$
dependence was assumed both for $H_1^{q\perp}$ and $D_1^q$ and
integrated over $|p_{h\perp}|$.}
\be
{{\rm d}\sigma\over {\rm
d}\cos\theta_2 {\rm d}\phi_1}\propto (1+\cos^2\theta_2)\cdot \left(1+
{6\over\pi}\left[{H_1^{\perp q}\over D_1^q}\right]^2 C_{TT}^{q\bar
q}{\sin^2\theta_2\over 1+\cos^2\theta_2}\cos(2\phi_1)\right) \, ,
\label{mulders}
\ee
where $\theta_2$ is the polar angle of the electron beam relative
to the second hadron momenta ${\bm p}_2$, and $\phi_1$ is
the azimuthal angle of the first hadron counted off the $({\bm
p}_2,\, {\bm e}^-)$-plane. This asymmetry was probed \cite{todd}
using the DELPHI data collection 91--95.  For the leading charged
particles (mostly pions) in each jet of two-jet events, summed
over $z$ and averaged over quark flavours (assuming
$H_1^{\perp}=\sum_H H_1^{\perp\, q/H}$ is flavour independent),
the most reliable preliminary value of the analyzing power
(obtained from the region $45^\circ<\theta_2<135^\circ$ with
small acceptance corrections) is found to be $\left|{\la
H_1^{\perp}\ra\over\la D_1\ra}\right| =(6.3\pm 2.0)\%$. However,
the larger "optimistic" value
\begin{equation}
\left|{\la H_1^{\perp}\ra\over\la D_1\ra}\right| =(12.5\pm 1.4)\%
\label{apower}
\end{equation}
(obtained from the whole acceptable region
$15^\circ<\theta_2<165^\circ$, see Fig. \ref{fig4}) is not
excluded, with smaller statistical but presumably large
systematic errors. This value, as it will be seen below, better fits
the description of the azimuthal asymmetries in SIDIS with
longitudinally polarized target.

\section{Chiral quark-soliton model prediction for
{\boldmath $h_1^a(x)$}}
\label{sect-hiQSM}

In order to make quantitative estimates for asymmetries we will
use for the transversity distribution function predictions from
the chiral quark-soliton model ($\chi$QSM)
\cite{Schweitzer:2001sr}. This model was derived from the
instanton model of the QCD vacuum \cite{Diakonov:2002fq} and
describes numerous nucleonic properties without adjustable
parameters to within $(10-30)\%$ accuracy \cite{Christov:1995vm}.
The field theoretic nature of the model allows to consistently
compute quark and antiquark distribution functions
\cite{Diakonov:1996sr} which agree with parameterizations
\cite{Gluck:1994uf} to within the same accuracy. This gives us a
certain confidence that the model also describes $h_1^a(x)$ with
a similar accuracy.

\begin{figure}[ht]
\begin{center}
\vspace{-10mm}
\includegraphics[width=.37\textwidth]{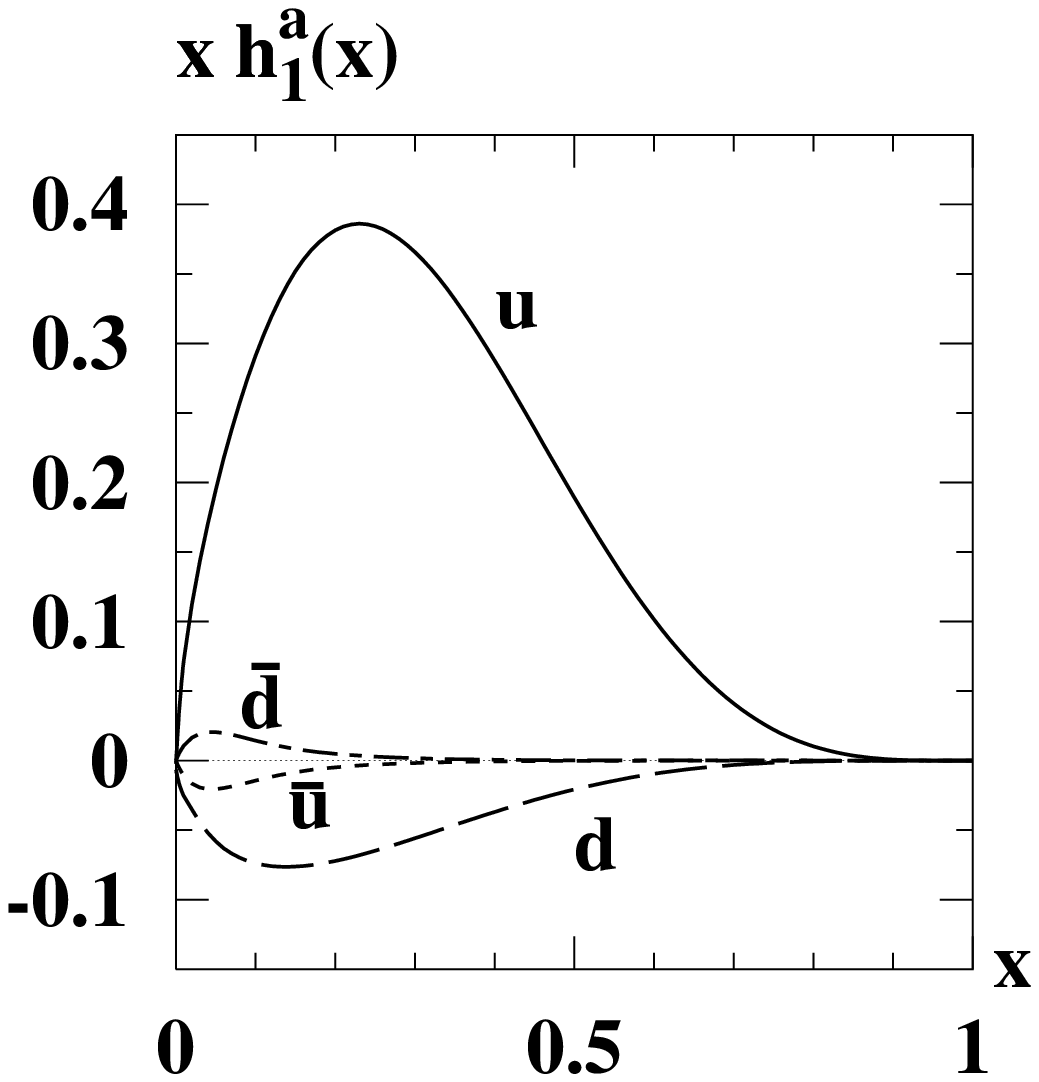}~a)
\includegraphics[width=.37\textwidth]{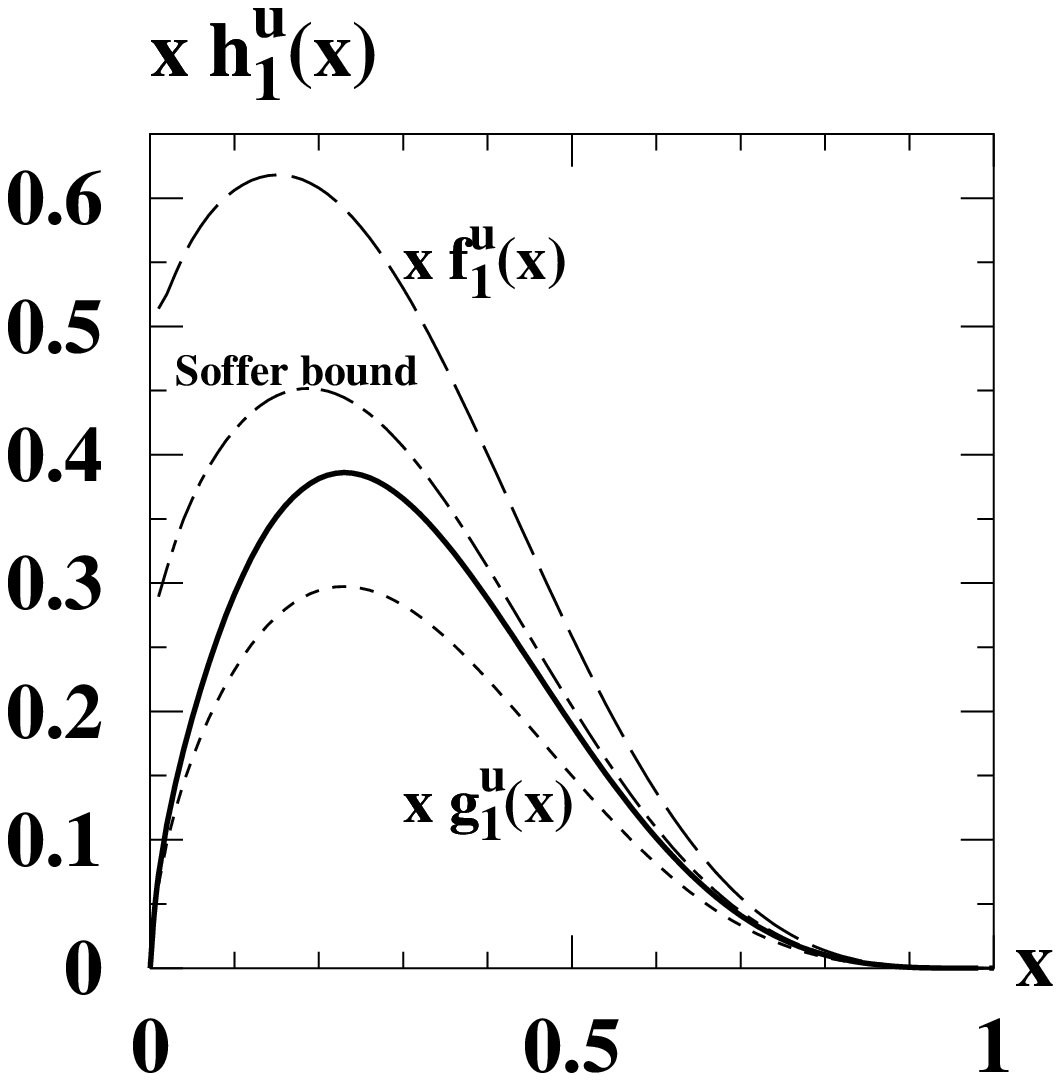}~b) 
\end{center}
\caption{{\bf a)}
    The transversity distribution function $h_1^a(x)$ vs.\  $x$
    from the $\chi$QSM \cite{Schweitzer:2001sr}.
    \newline {\bf b)}
    Comparison of $h_1^u(x)$ from the $\chi$QSM (solid) to
    $f_1^u(x)$ (dashed) and $g_1^u(x)$ (dotted) and the Soffer
    bound $(f_1^u+g_1^u)(x)/2$ (dashed-dotted line) taken or
    constructed from the parameterizations of
    \cite{Gluck:1994uf}.}
    \label{fig-h1}
\end{figure}

In the $\chi$QSM we observe the hierarchy
$h_1^u(x)\gg|h_1^d(x)|\gg|h_1^{\bar u}(x)|$, and an interesting
"maximal sea quark flavour asymmetry" $h_1^{\bar d}(x)\approx
-h_1^{\bar u}(x)>0$. In Fig.~\ref{fig-h1}a we show the $\chi$QSM
prediction for $h_1^a(x)$ from Ref.~\cite{Schweitzer:2001sr}
LO-evolved from the low scale of the model of about
$\mu_0^2=(0.6\,{\rm GeV})^2$ to the scale $Q^2=16\,{\rm GeV}^2$.
In order to gain some more intuition on the predictions we
compare in Fig.~\ref{fig-h1}b the dominating distribution
function $h_1^u(x)$ from the $\chi$QSM to $f_1^u(x)$ and
$g_1^u(x)$ from the parameterizations of
Ref.~\cite{Gluck:1994uf}. It is remarkable that the Soffer
inequality \cite{Soffer:1995ww} $|h_1^u(x)| \le
(f_1^u+g_1^u)(x)/2$ is nearly saturated -- in particular in the
large-$x$ region. (The Soffer bound in Fig.~\ref{fig-h1}b is
constructed from $f_1^u(x)$ and $g_1^u(x)$ taken at $Q^2=16\,{\rm
GeV}^2$ from \cite{Gluck:1994uf} in comparison with $h_1^u(x)$.)
For the unpolarized distribution function $f_1^a(x)$ we use the
LO parameterization from Ref.~\cite{Gluck:1994uf}.

\section{Comparison with longitudinally polarized data }
\label{sect-longpol}

In the HERMES experiments
\cite{Airapetian:1999tv,Airapetian:2001eg,Airapetian:2002mf} the
$x$- and $z$-dependence of asymmetries (\ref{asim}) for pions and
kaons with $W=\sin\phi$ or $\sin2\phi$ from longitudinally
(relative to the non-polarized lepton beam) polarized proton and
deuterium target were measured. As it is seen from (\ref{pdfpff})
only the second term of $d\sigma_{0L}$ and the first (Collins)
and the third (Sivers) ones\footnote{\label{chiral-t-odd}
Actually, our approach would imply the vanishing of the Sivers
effect. This is in agreement with the $\chi$QSM. However, this
cannot be taken literally as a prediction for the following
reason. The $\chi$QSM was derived from the instanton vacuum model
as the leading order in terms of the instanton packing fraction
$\frac {\rho}{R}\sim \frac{1}{3}$ ($\rho$ and $R$ are
respectively the average size and separation of instantons in
Euclidean space time). In this order the T-odd PDF's like
$f_{1T}^\perp$ and $h_{1}^\perp$ vanish \cite{Pobylitsa:2002fr}.
In higher orders the T-odd PDF's can be well non-zero and all one
can conclude at this stage is that the T-odd PDF's are suppressed
with respect to the T-even. However, considering that
$H_1^\perp(z)$ is much smaller than $D_1(z)$, cf.\
Eq.~(\ref{apower}), it is questionable whether this suppression
could be sufficient such that in physical cross sections the
Collins effect $\propto h_1^a(x)H_1^\perp(z)$ is dominant over
the Sivers  effect $\propto f_{1T}^\perp(x)D_1(z)$. (For an
estimation of this suppression see
\cite{Efremov:2003tf,Schweitzer:2003yr}.)
}
of $d\sigma_{0T}$ with factor (\ref{PT}) do contribute to the
numerator of (\ref{asim}) for $W=\sin\phi$ with negative sign
($\phi_S=\pi$) and only the third one of the $d\sigma_{0L}$
contributes for $W=\sin2\phi$. This gives for 
\bea
\label{AUL-sinPhi}
A_{0L,P\mbox{\tiny or}D}^{\sin\phi}(x,z,h) &=& P_{\! L}(x)\;
\frac{\sum_a e_a^2\, x h_L^{a/P\mbox{\tiny or}D}(x)\,H_1^{\perp a}(z)}
{\sum_{a}^h e_{a}^2\, f_1^{a/P\mbox{\tiny
or}D}(x)\,D_1^{a}(z)\,}\\
\nonumber
&-& P_{\! T}(x)\;
\frac{\sum_a e_a^2\, h_1^{a/P\mbox{\tiny or}D}(x)\,H_1^{\perp a}(z)}
{\sum_{a}^h e_{a}^2\,f_1^{a/P\mbox{\tiny
or}D}(x)\,D_1^{a}(z)\,}\\
\label{AUL-sin2Phi}
A_{0L,P\mbox{\tiny or}D}^{\sin2\phi}(x,z,h) &=&
P_{\! 1}(x)\;
\frac{\sum_a e_a^2\, h_{1L}^{\perp(1),a/P\mbox{\tiny or}D}(x)\,
H_1^{\perp a}(z)}{\sum_{a}^h e_{a}^2\, f_1^{a/P\mbox{\tiny
or}D}(x)\,D_1^{a}(z)}\,,
\eea
where $P_{\!L}(x)$ and $P_{\!T}(x)$ are known factors of order
${\cal O}(M/Q)$ and $P_{\! 1}$ of order ${\cal O}(1)$ depending
on average transverse momenta of partons inside hadron and
hadrons inside parton. The functions $h_L^{a}(x)$ and
$h_{1L}^{\perp(1),a}(x)$ are expressed through transversity
$h_1^{a}(x)$ by the relation (\ref{wwform}).

For $H_1^{\perp a}$ and $D_1^{a}$ a strong suppression of the
unfavoured with respect to the favoured fragmentation has been
assumed. From charge conjugation and isospin symmetry one has
then
\begin{equation}
\label{H1-favor}
H_1^{\perp\rm fav} \equiv H_1^{\perp
u/\pi^+}\!\! = H_1^{\perp d/\pi^-}\!\! = 2H_1^{\perp
u/\pi^0}\dots\gg H_1^{\perp d/\pi^+}\!\! = H_1^{\perp
u/\pi^-}\dots \equiv H_1^{\perp\rm unf}\,.
\end{equation}

\begin{figure}[h!]
\vspace{-3mm}
\begin{center}
\includegraphics[width=4.0cm]{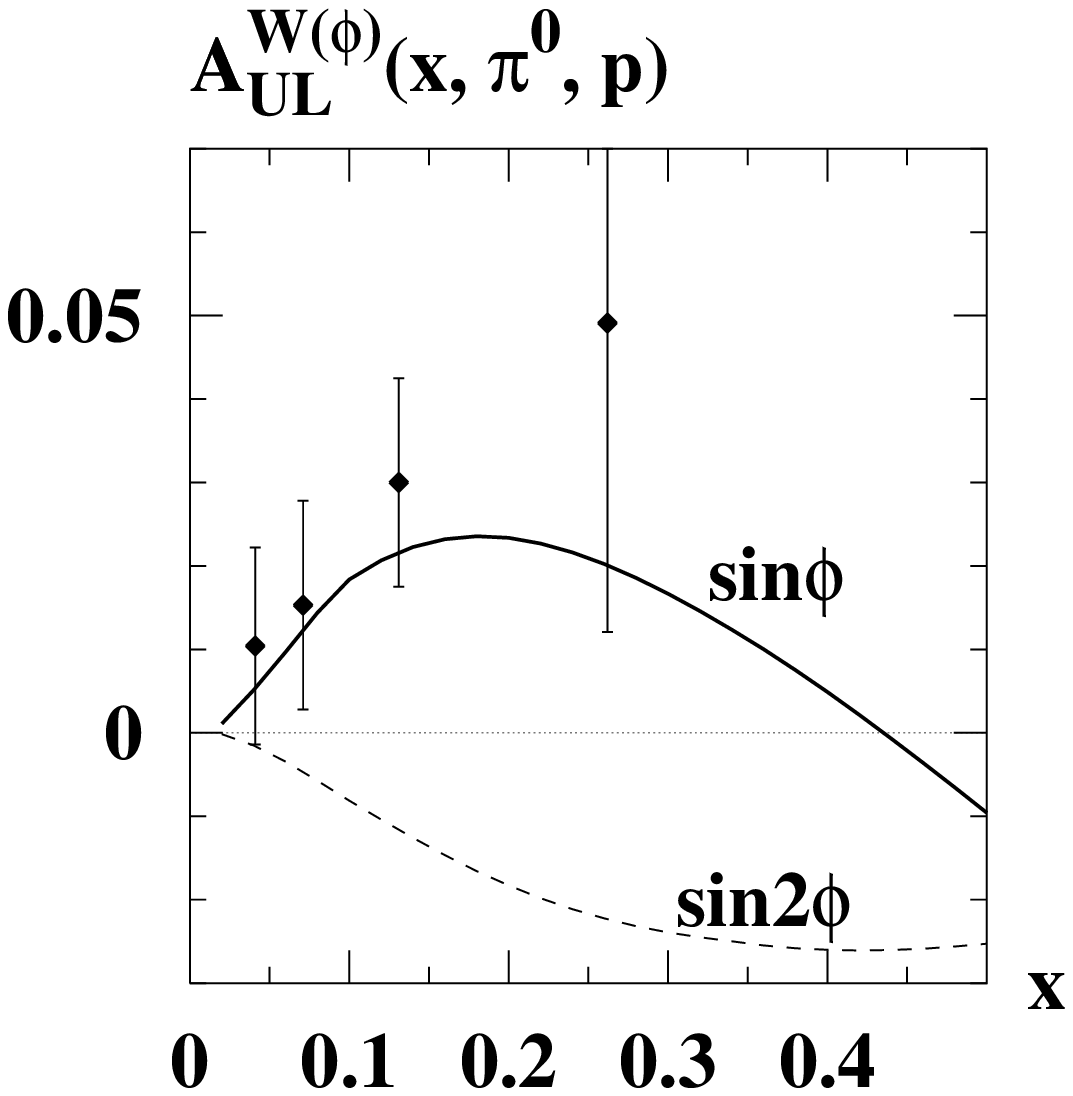}
\includegraphics[width=4.0cm]{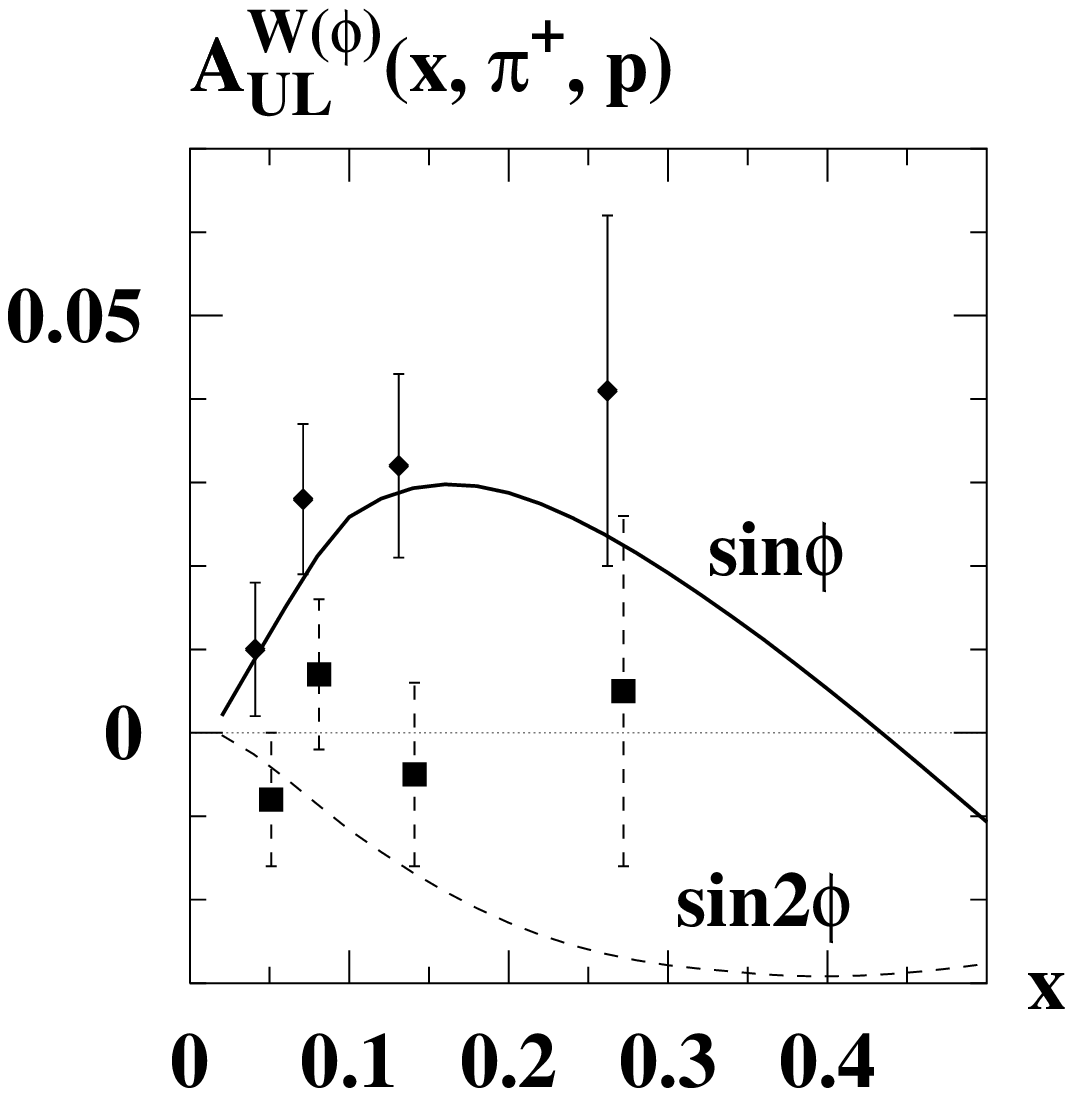}
\includegraphics[width=4.0cm]{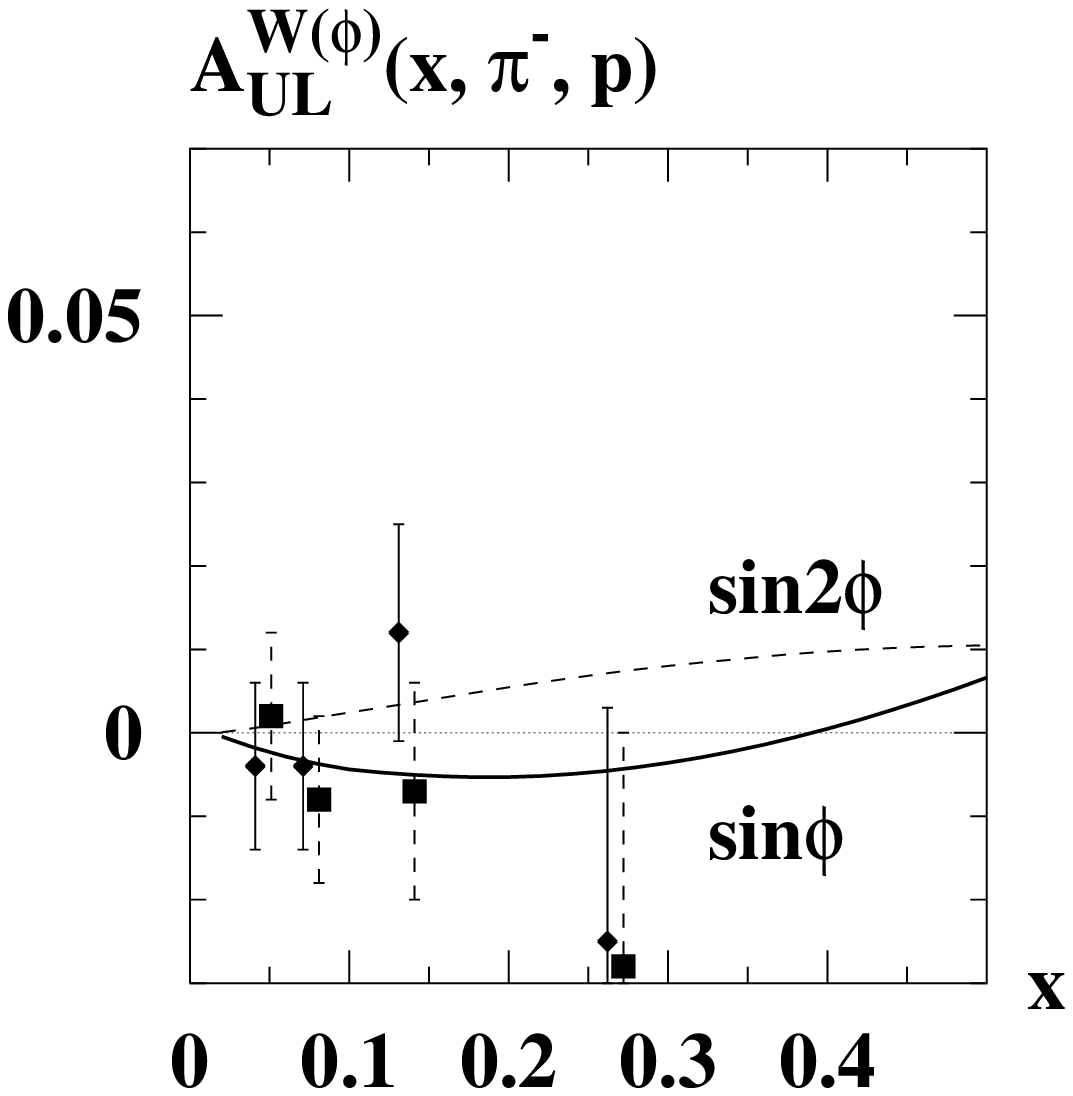}
\end{center}
\vspace{-5mm}
\caption{ Azimuthal asymmetries
$A_{0L}^{W(\phi)}(x,\pi)$ weighted by $W(\phi)=\sin\phi$ (solid
line) and $\sin 2\phi$ (dashed line) for the production of
$\pi^0$, $\pi^+$ and $\pi^-$ as function of $x$. The experimental
data are from the Refs. \cite{Airapetian:1999tv,Airapetian:2001eg}.
Rhombus (squares) denote the data for $A_{0L}^{\sin\phi}$
($\;A_{0L}^{\sin2\phi}$). }
\label{AUL-prot}
\end{figure}

So, using the DELPHI result\footnote{
We assume a  weak scale
dependence of the analyzing power (\ref{apower}).
}
Eq.(\ref{apower}), $\chi$QSM for $h_1^a(x)$, the relation
(\ref{wwform}) for the $h_L$ and the parameterization from Ref.
\cite{Gluck:1994uf} for $f_1^a(x)$, both LO-evolved to the
average scale $Q_{\rm av}^2=4\,{\rm GeV}^2$ characteristic for
HERMES we obtain for the proton target 
Fig.~\ref{AUL-prot}.
\begin{figure}[t]
\begin{minipage}{60mm}
\vspace{-2mm}
\includegraphics[width=1\textwidth]{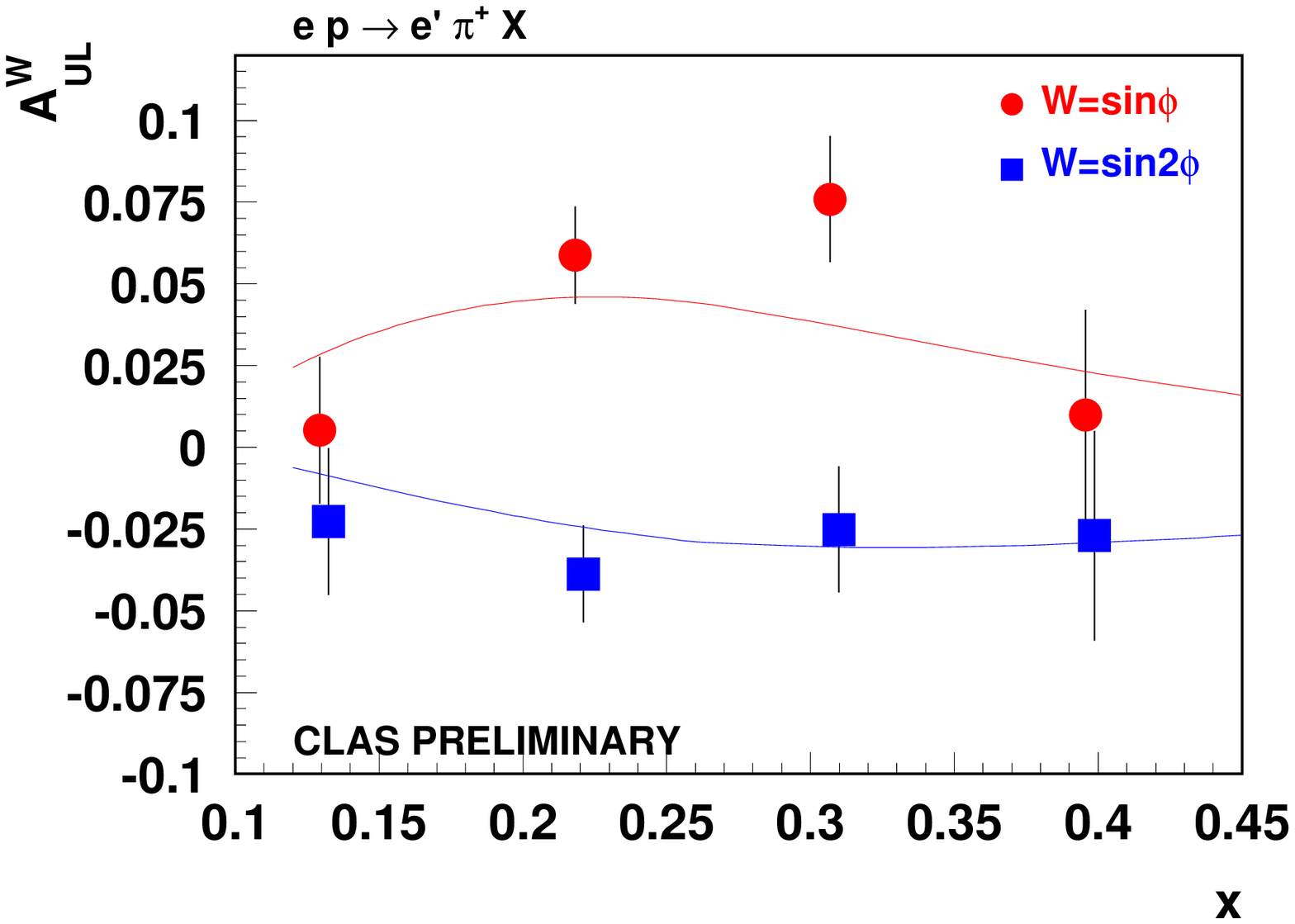}\\[-8mm]
\caption{Our predictions (solid curves) for azimuthal asymmetries
$A_{0L,D}^{\sin2\phi}$ and $A_{0L,D}^{\sin\phi}$ vs. $x$ in
comparison with CLAS data \cite{Avakian:2002qp}.}
\label{AUL-sin2phi}
\end{minipage}
\hfill
\begin{minipage}{60mm}
\vspace{-4mm}
\includegraphics[width=.9\textwidth,height=.24\textheight]
{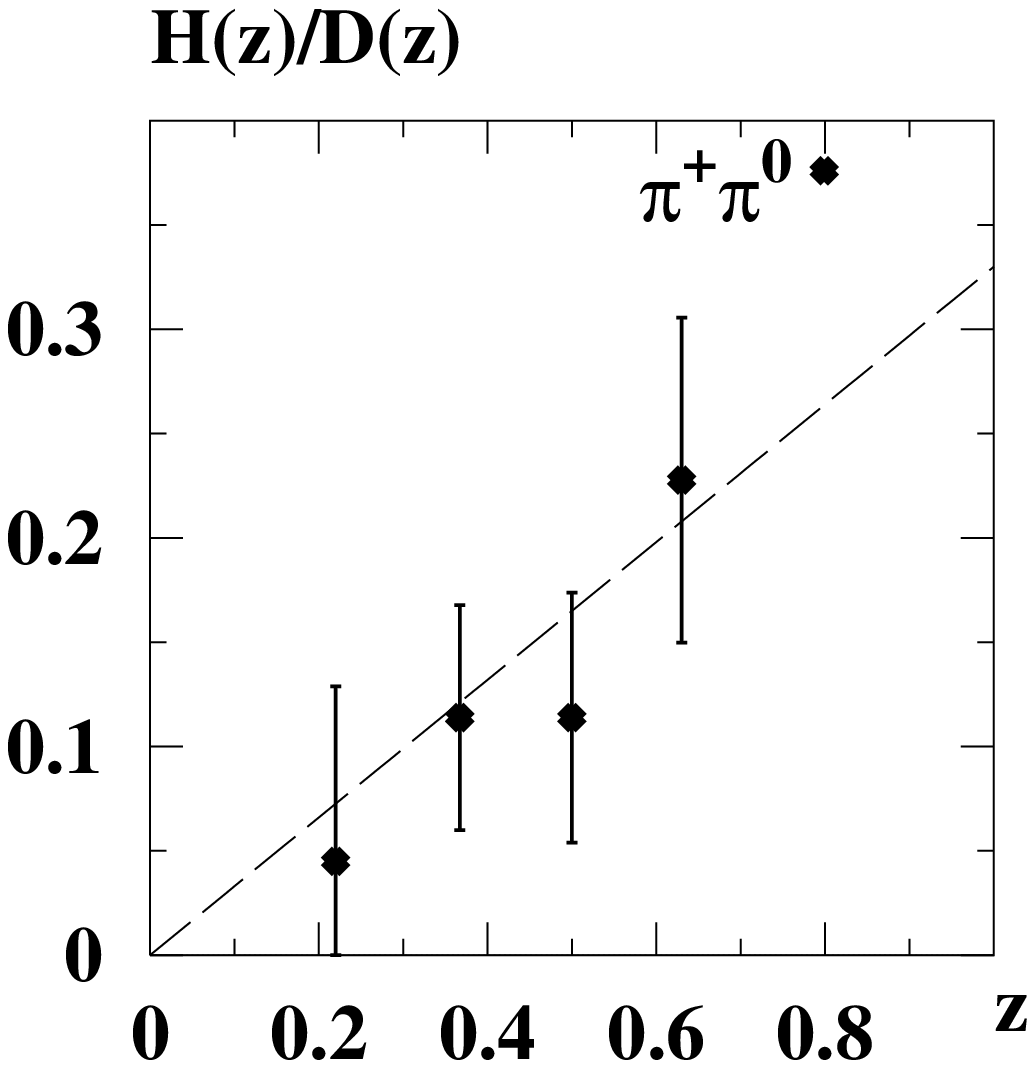}\\[-5mm]
\caption{
$H_1^\perp/D_1$ vs. $z$, as extracted from HERMES data
\cite{Airapetian:1999tv,Airapetian:2001eg} for $\pi^+$ and
$\pi^0$ production combined.}
\label{fig-H(z)}
\end{minipage}
\end{figure}
\begin{wrapfigure}[30]{HR}{5.3cm}
\begin{flushright}
\includegraphics[width=.38\textwidth,height=.53\textheight]
{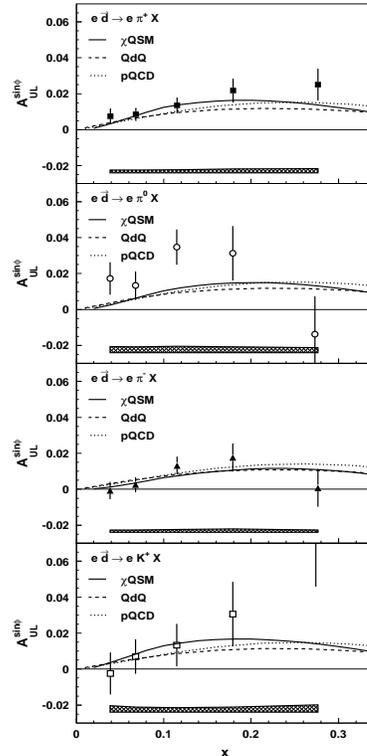}\\[-7mm]
\begin{minipage}{4.3cm}
\caption{Our predictions (solid curves) for $A_{0L,D}^{\sin\phi}(x,h)$ vs.
$x$ from a longitudinally polarized deuteron target in comparison
with the HERMES data \cite{Airapetian:2002mf}. ${}\,{}$}
\end{minipage}
\end{flushright}
\label{AUL-deut-sinphi}
\end{wrapfigure}

Fig. \ref{AUL-sin2phi} shows also our predictions for
$A_{0L,P}^{\sin\phi}(x,\pi)$ and $A_{0L,P}^{\sin2\phi}(x,\pi)$ in
comparison with CLAS data \cite{Avakian:2002qp} for a 5.7 GeV
beam.

We conclude that the azimuthal asymmetries obtained with the
$\chi$QSM prediction for $h_1^a(x)$ \cite{Schweitzer:2001sr}
combined with the ``optimistic'' DELPHI result (\ref{apower}) for
the analyzing power are consistent with experiment {\sl with no
fit parameter}.

It is interesting to note that the negative sign of the
transversal contribution in (\ref{AUL-sinPhi}) leads to a change
of sign of the $A^{\sin\phi}_{0L}$ asymmetries for $x>0.4$. This
is due to a harder behaviour $h_1(x)$ with respect to $h_L(x)$
followed from Eq. \ref{wwform} . It should be noted that the
prediction of $A_{0L}^{\sin\phi}(x,\pi)=0$ at $x\simeq(0.4-0.5)$
is sensitive to the approximation of favoured flavour
fragmentation (\ref{H1-favor}). In principle one could conclude
from data, how well this approximation works. However, the upper
$x$-cut is $x<0.4$ in the HERMES experiment
\cite{Airapetian:1999tv,Airapetian:2001eg}.

Now let us reverse the logic, and using z-dependence of the
HERMES results for the $\pi^0$ and $\pi^+$ azimuthal asymmetries
try to estimate $H_1^\perp(z)/D_1(z)$. For that we use the
$\chi$QSM prediction for $h_1^a(x)$ that will introduce a model
dependence of order (10 - 20)\%. The combined result is shown in
Fig. \ref{fig-H(z)}.  The data can be described by a
linear fit
\be
H_1^\perp(z)=(0.33 \pm  0.06)zD_1(z)
\label{apower-vs-z}
\ee
with average ${\la H_1^\perp\ra}/{\la D_1\ra}=(13.8\pm 2.8)\%$
which is in good agreement with DELPHI result Eq.(\ref{apower}).
The errors are the statistical errors of the HERMES data. It is
interesting to note that numerically the behaviour
(\ref{apower-vs-z}) is close to those calculated from chirally
invariant Manohar-Georgi model \cite{Bacchetta:2002es},
$H_1^\perp(z)\approx 0.63z^2D_1(z)$.

Believing that such an acceptable description  of the proton data
is not occasional we made the predictions \cite{Efremov:2001ia}
for $A_{0L}$ asymmetries for pions and kaons at longitudinally
polarized deuteron target which were being measured at that time
by the HERMES collaboration.

The main question was, however, how large is the analyzing power
for kaons? We know that the unpolarized kaon fragmentation
function $D_1^K(z)$ is roughly five times smaller than the
unpolarized pion one. Is also $H_1^{\perp\,K}(z)$ five times
smaller than $H_1^{\perp\,\pi}(z)$? {The reason is that in 
the chiral limit $D_1^\pi=D_1^K$ and $H_1^{\perp\pi}=H_1^{\perp
K}$. The naive expectation is that the 'way off chiral limit to
real world' proceeds analogously as for spin-dependent
quantities, $H_1^\perp$, as for spin-independent one, $D_1$.} If
we assume this, i.e. if
\be
\frac{\la H_1^{\perp\,  K}\ra}{\la D_1^K  \ra} \simeq
\frac{\la H_1^{\perp\,\pi}\ra}{\la D_1^\pi\ra}
\label{apower-kaon}
\ee
holds, we obtain -- with the central value of $\la
H_1^\perp\ra/\la D_1\ra$ in Eq.(\ref{apower}) -- azimuthal
asymmetries for $K^+$ and $K^0$ as large as for pions. The
results of our predictions \cite{Efremov:2001ia} (solid curves)
in comparison with the published HERMES data
\cite{Airapetian:2002mf} are presented at Fig.
\ref{AUL-deut-sinphi}. Again no fit parameters were used in
distinction with other models at Fig. \ref{AUL-deut-sinphi}. The
asymmetries for $\bar K^0$ and $K^-$ are close to zero in our
approach.

\begin{wrapfigure}[14]{RH}{5.0cm}
\vspace{-13mm}
\begin{flushright}
\includegraphics[width=.35\textwidth]
{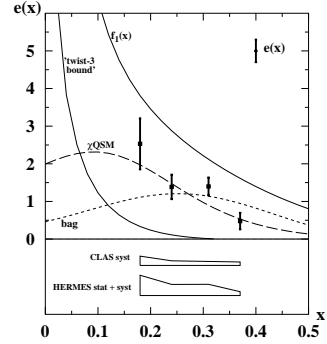}\\[-5mm]
\begin{minipage}{4.85cm}
\caption{ The flavour combination
$e(x)\!=\!(e^u\!+\!\frac{1}{4}e^{\bar d})(x)$, with error bars
due to statistical error of CLAS data, vs. $x$ at $\la
Q^2\ra\!=\! 1.5\,{\rm GeV}^2$. For comparison the twist-3 bound,
$f_1^u(x)$, bag and $\chi$QSM models predictions are shown. }
\label{fig-eofx}
\end{minipage}
\end{flushright}
\end{wrapfigure}

\section{Extraction of {$\mathbf e(x)$} from $\mathbf{A_{L0}}$}

Very recently the $\sin\phi$ asymmetry of $\pi^+$ produced by
scattering of polarized electrons off unpolarized protons was
published by CLAS collaboration \cite{Avakian:2003pk} and
preliminary data were reported by HERMES collaboration
\cite{Avetisyan:2004uz}. This asymmetry is interesting since it
allows to access the unknown twist-3 structure functions $e^a(x)$
(see $d\sigma_{L0}$ in (\ref{pdfpff})) that are connected with the
nucleon $\sigma$-term \cite{Schweitzer:2003sb}:\\
\begin{minipage}{73mm}
\be
\label{sigma}
    \int_0^1\di x\sum_ae^a(x)=\frac{2\sigma}{m_u+m_d}
    \approx 10 \; .
\ee
\end{minipage}\\[3mm]

The asymmetry is given by\footnote{The term $\propto \sum_a
e_a^2h_1^{\perp a}(x)\la E^{a/\pi}\ra$ in numerator was
disregarded since $h_1^{\perp a}$ are suppressed in $\chi$QSM
(see footnote \ref{chiral-t-odd}.)}

\begin{minipage}{68mm}
\be
\hspace{-3mm}
    A_{L0}^{\sin\phi}(x)\propto \frac{M}{Q} \;
    \frac{\sum_a e_a^2e^a(x)\la H^{\perp a/\pi}_1\ra}
    {\sum_a e_a^2 f^a_1(x)\la D_1^{a/\pi}\ra}  \;.
\label{ALU}
\ee
\end{minipage}\\[3mm]

Disregarding unfavoured fragmentation and using the Collins
analysing power extracted from HERMES in (\ref{apower-vs-z}) that
yields for $z$-cuts of CLAS ${\la H^{\perp\pi}_1\ra}/{\la
D_1^\pi\ra}=0.20 \pm 0.04$, we can extract \cite{Efremov:2002ut}
$e^u(x)+\frac14e^{\bar d}(x)$. The result is presented in Fig.
\ref{fig-eofx}. For comparison the twist-3 lower
bound\footnote{\label{footnote-Soffer}Let us stress that strictly
speaking this inequality could be justified only if the ``twist-2
Soffer inequality'' $2|h_1^a(x)|\le(f_1^a+g_1^a)(x)$ of
Ref.~\cite{Soffer:1995ww} were saturated \cite{Efremov:2002qh}.
In the following we will refer to this relation as ``twist-3
lower bound'' keeping in mind that it does not need to hold in
general. In Ref.~\cite{Lu:1996ae} a bound based on the positivity
of the hadronic tensor and the Callan-Gross relation (and
formulated in terms of structure functions) was discussed.
},
$e^a(x)\ge 2|g_T^a(x)|-h_L^a(x)$ Ref.~\cite{Soffer:1995ww}, and
the unpolarized distribution function $f_1^u(x)$ are plotted. The
predictions of $\chi$QSM \cite{Schweitzer:2003uy} and bag model
\cite{Signal:1997ct} are shown also. It seems that $e^a(x)$ is
not small, but not sizeable enough to explain the large number in
Eq.(\ref{sigma}). However, QCD equations of motion imply a
$\delta$-function at $x=0$ in $e^a(x)$, see \cite{Efremov:2002qh}
and references therein, and it is the $\delta(x)$ contribution
which entirely saturates the sum rule, giving rise to the large
number in (\ref{sigma}). Remarkably, also in the $\chi$QSM
$e^a(x)$ contains a $\delta(x)$ \cite{Schweitzer:2003uy}.

\section{SIDIS with transverse target polarization}
\label{sect-transpol}

In the HERMES and COMPASS experiments the cross sections
$\sigma_N^{\uparrow\downarrow}$ for the process
$lN^{\uparrow\downarrow}\rightarrow l'h X$ is measured, where
$N^{\uparrow\downarrow}$ denotes the target state transversely
polarized with respect to the beam (see Fig.~\ref{kinsidis}).

The component of the target polarization vector which is {\sl
transverse relative to the hard photon} is characterized by the
angle $\Theta_S$, see Fig.~\ref{kinsidis}, given by
\be
\label{spin-projection}
\hspace{-8mm}
\sin\Theta_S
=\cos\theta_\gamma\sqrt{1+{\rm tan}^2\theta_\gamma\sin^2\phi_{S'}}
\approx \cos\theta_\gamma \approx 1\;,
\ee
where $\phi^\prime_S$ is the azimuthal angle of the target
polarization direction about the lepton beam direction relative
to the scattering plane. As it is seen from (\ref{pdfpff}) only
the first (Collins) term of $d\sigma_{0T}$ with factor
(\ref{spin-projection}) does contribute to the numerator of
(\ref{asim}) for $W=\sin(\phi+\phi_S)$. This gives for the
Collins asymmetry
\be
A_{0T}^{\sin(\phi+\phi_s)}(x,z,h) = B_T(x)\;
\frac{\sum_a e_a^2\,x\, h_1^a(x)\,H_1^{\perp a}(z)}
{\sum_b e_b^2\,x\, f_1^b(x)\,D_1^b(z)\,} \;,
\label{AUT-without-kT-fin}
\ee
with the known factor $B_{\!T}(x)$  of ${\cal O}(1)$ depending on
average transverse momenta.

\begin{figure}[t]
\begin{center}
\includegraphics[width=.30\textwidth]
{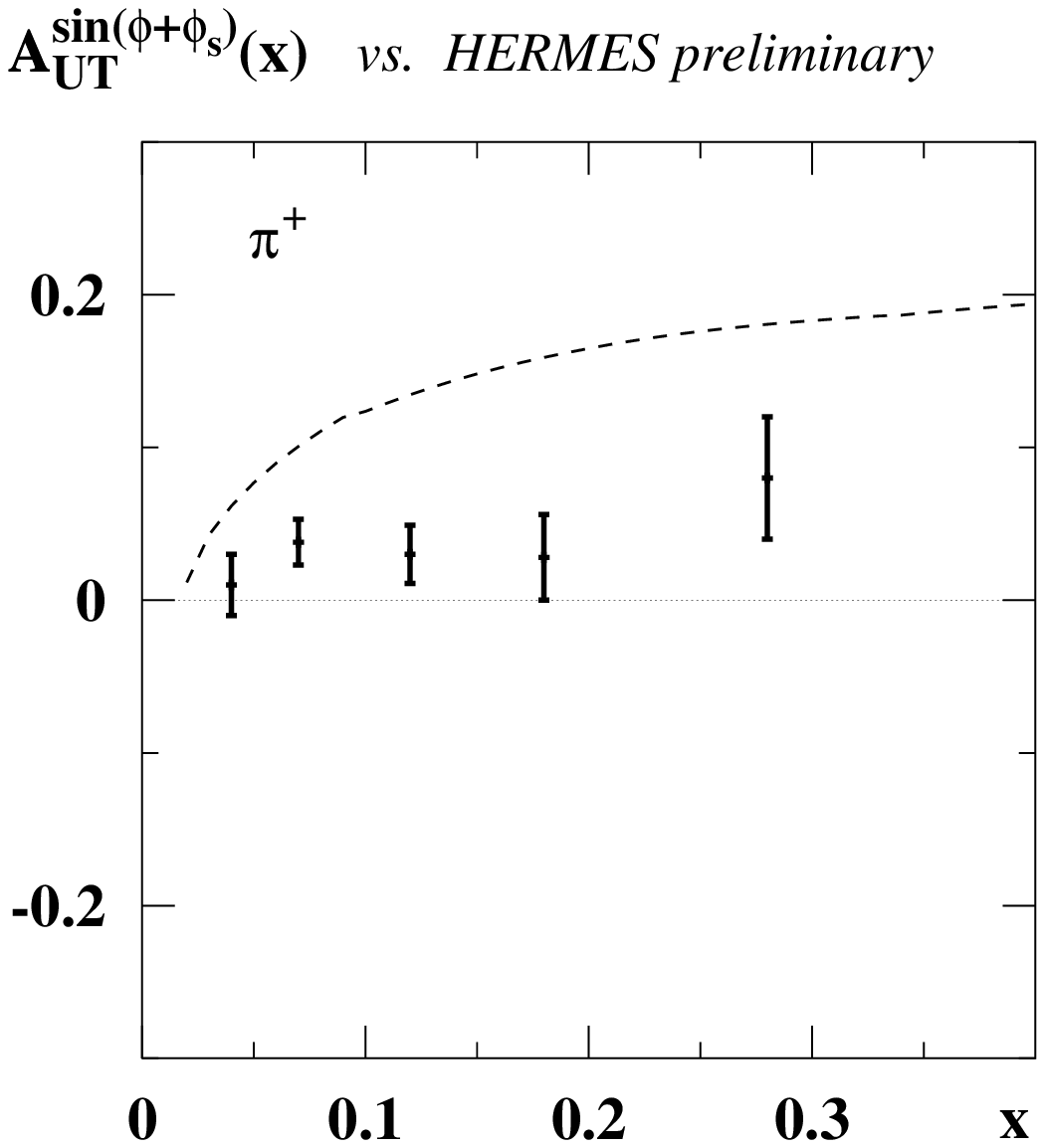}
\includegraphics[width=.30\textwidth]
{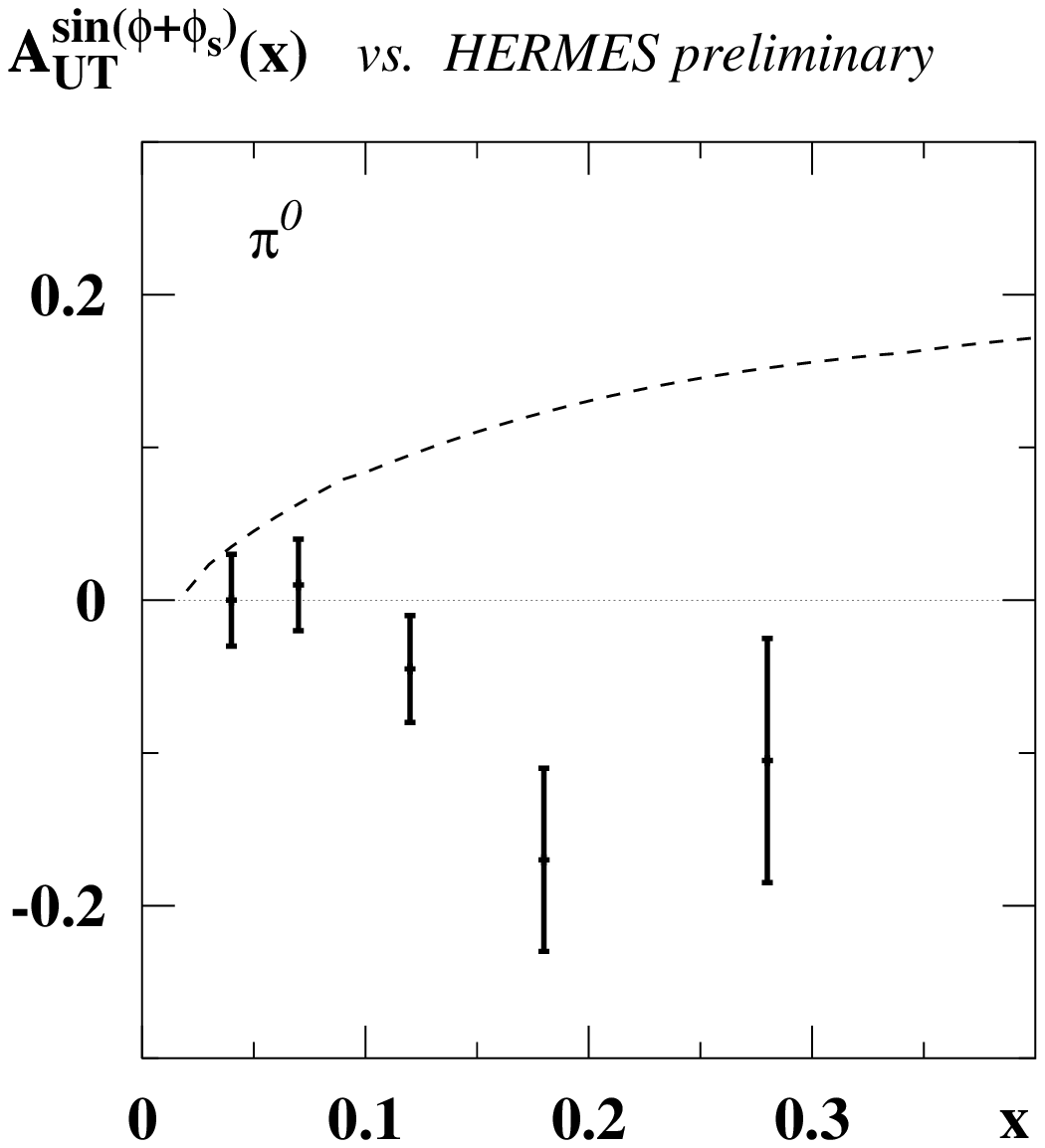}
\includegraphics[width=.30\textwidth]
{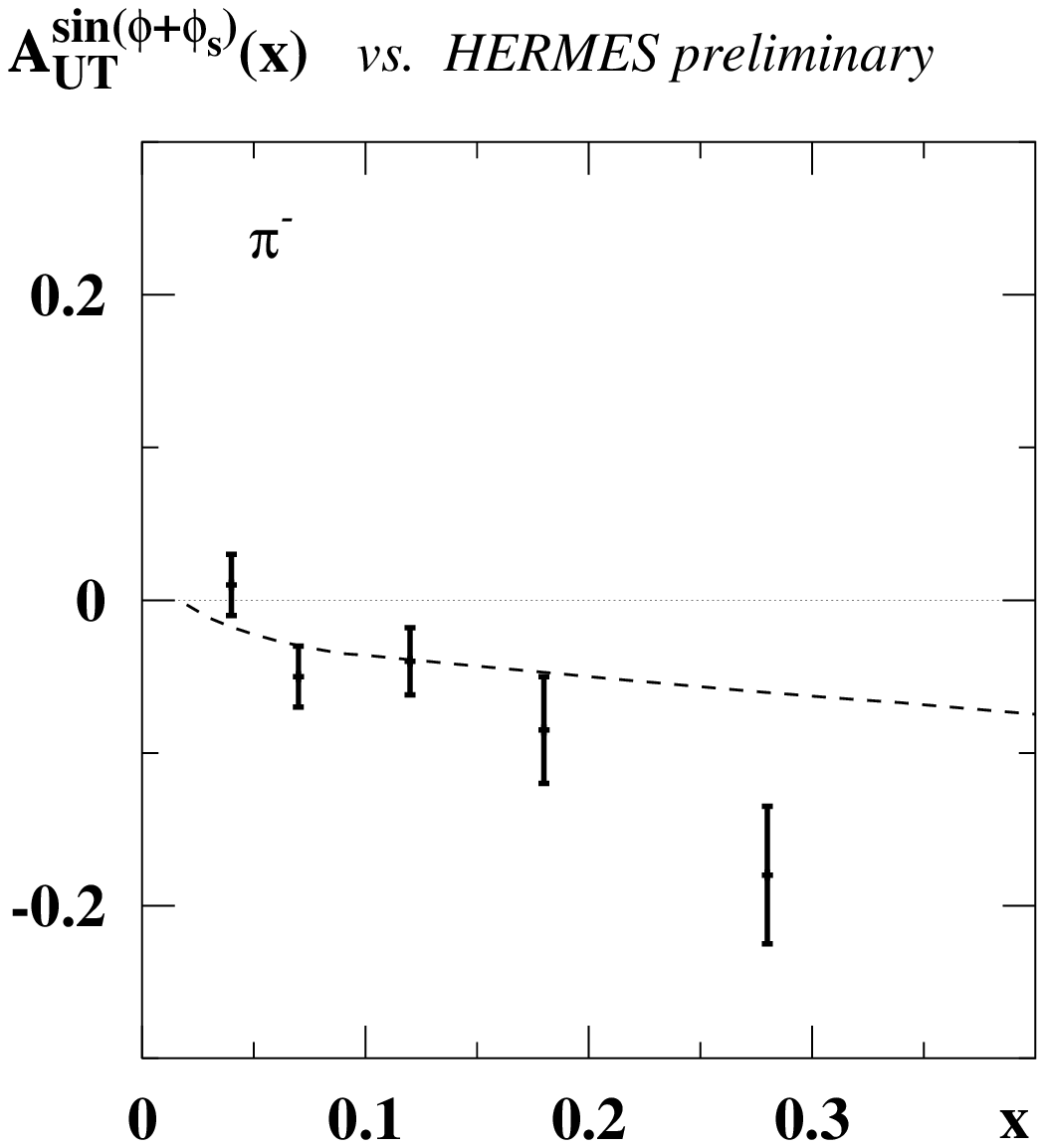}
\end{center}
\vspace{-0.3cm} \caption[*]{ The Collins effect transverse target
SSA $A_{0T}^{\sin(\phi+\phi_s)}$ in the production of $\pi^+$,
$\pi^0$ and $\pi^-$ from a proton target. {\sl Preliminary} data
are from \cite{AUT-HERMES}, theoretical curves from
\cite{Efremov:2003eq}.}
\label{AUT-col-HERMES}
\vspace{-0.5cm}
\end{figure}

Based on our understanding of the longitudinally polarized target
asymmetry, predictions for the Collins effect for transversally
polarized target asymmetry $A_{0T}^{\sin(\phi+\phi_s)}$ were
made\footnote{Asymmetries of similar magnitude as for HERMES are
also predicted \cite{Efremov:2003eq} for the running COMPASS
experiment with transversally and longitudinally polarized
targets.
}
\cite{Efremov:2003eq} (see Fig. \ref{AUT-col-HERMES}). Of course,
since the theoretical description of the power suppressed
(``twist-3'')  longitudinal asymmetry is involved and we made
simplifications, which are difficult to control, one cannot
expect that we accurately predict the overall magnitude of the
effect. However, one could have a certain confidence that at
least {\sl the sign and the shape} of
$A_{0T}^{\sin(\phi+\phi_s)}(x)$ is described satisfyingly since
it is dictated by the model prediction for $h_1^a(x)$
\cite{Schweitzer:2001sr} and the approximation of favoured
flavour fragmentation only. As can be seen in Fig.
\ref{AUT-col-HERMES} our results \cite{Efremov:2003eq} do not
even describe the sign for $\pi^0$ of the preliminary HERMES data
\cite{AUT-HERMES}! Why not?

Apparently some assumption(s) made must be incorrect. The first
suspicion is favoured fragmentation approximation
(\ref{H1-favor}). First of all, pay attention to the negative and
large $A_{0T}^{\sin(\phi+\phi_s)}(\pi^0)$. With the unfavoured
fragmentation taken into account one has from charge conjugation
and isospin invariance (\ref{H1-favor})

\begin{equation}
\label{pi0-frag}
H_1^{\perp a/\pi^0}=
\frac{1}{2}(H_1^{\perp a/\pi^+}+H_1^{\perp a/\pi^-})
=\frac{1}{2}(H_1^{\perp\rm fav}+ H_1^{\perp\rm unf})\,.
\end{equation}
Then
\be
\label{AUT-unf-pi0}
\underbrace{A_{0T}^{\sin(\phi+\phi_s)}(\pi^0)}_{<0\;\rm
in\;experiment} \propto \underbrace{{\textstyle\sum_a}e_a^2
h_1^a(x)}_{>0\;\rm in\; models} \; \la H_1^{\perp\rm
fav}+H_1^{\perp\rm unf}\ra \;\;\; \Longrightarrow \;\;\; \la
H_1^{\perp\rm fav}+H_1^{\perp\rm unf}\ra <0 \;.
\ee

In order to explain the asymmetry for charged pions the option
$H_1^{\perp \rm fav}<0$ can be ruled out, unless
$(4h_1^u+h_1^{\bar d})<(h_1^d+4h_1^{\bar u})$ which would
contradict any model. Then, with the option $H_1^{\perp \rm
fav}>0$, we can draw two interesting conclusions from the
observation in Eq.~(\ref{AUT-unf-pi0}). Firstly, $H_1^{\perp \rm
unf}$ should have opposite sign with respect to $H_1^{\perp \rm
fav}$. This could have a natural explanation in string models, in
particular for the HERMES kinematics \cite{AUT-HERMES} with low
particle multiplicity jets. Secondly, the absolute value of
$H_1^{\perp \rm unf}$ has to be larger than the absolute value of
$H_1^{\perp \rm fav}$ which, if confirmed, will be more difficult
to understand.

Concerning the large value of $A_{0T}^{\sin(\phi+\phi_s)}(\pi^0)$
notice that it is approximately of the same order as
$A_{0T}^{\sin(\phi+\phi_s)}(\pi^-)$. Meanwhile from the
factorization of $x$ and $z$ dependence of polarized and
unpolarized SIDIS cross sections and from the relation
(\ref{pi0-frag}) one can write for {\sl any} spin asymmetry
\bea
A(\pi^0)&=&\frac{\sigma(\pi^+)}{\sigma(\pi^+)+\sigma(\pi^-)}\,A(\pi^+)
+\frac{\sigma(\pi^-)}{\sigma(\pi^+)+\sigma(\pi^-)}\,A(\pi^-)\\
\nonumber
&=&A(\pi^-)+a[A(\pi^+)-A(\pi^-)],
\eea
where $\sigma(\pi)$ is unpolarized SIDIS cross section. If
$A(\pi^0)\approx A(\pi^-)$ one  concludes
$a=\frac{\sigma(\pi^+)}{\sigma(\pi^+)+\sigma(\pi^-)}\approx 0$
that is nonsense! This leads to conclusion that the factorization
of $x$ and $z$ dependence for the transversally polarized SIDIS
cross sections is under suspicion and should be carefully
checked. Regrettably the statistical errors are rather large,
especially for $A_{0T}(\pi^0)$. Probably due to this reason the
final publication \cite{unknown:2004tw} does not contain
$A_{0T}(\pi^0)$. But even without $\pi^0$ the question stays why
$A_{0T}(\pi^+)$ is so small and $|A_{0T}(\pi^-)|$ is so large
since unfavoured $\la H_1^{\perp\rm unf}\ra$ gives opposite sign
contributions to $\pi^+$ and $\pi^-$ asymmetries the latter about
order of magnitude larger.

\begin{figure}[t!]
\begin{center}
\includegraphics[width=.30\textwidth]
{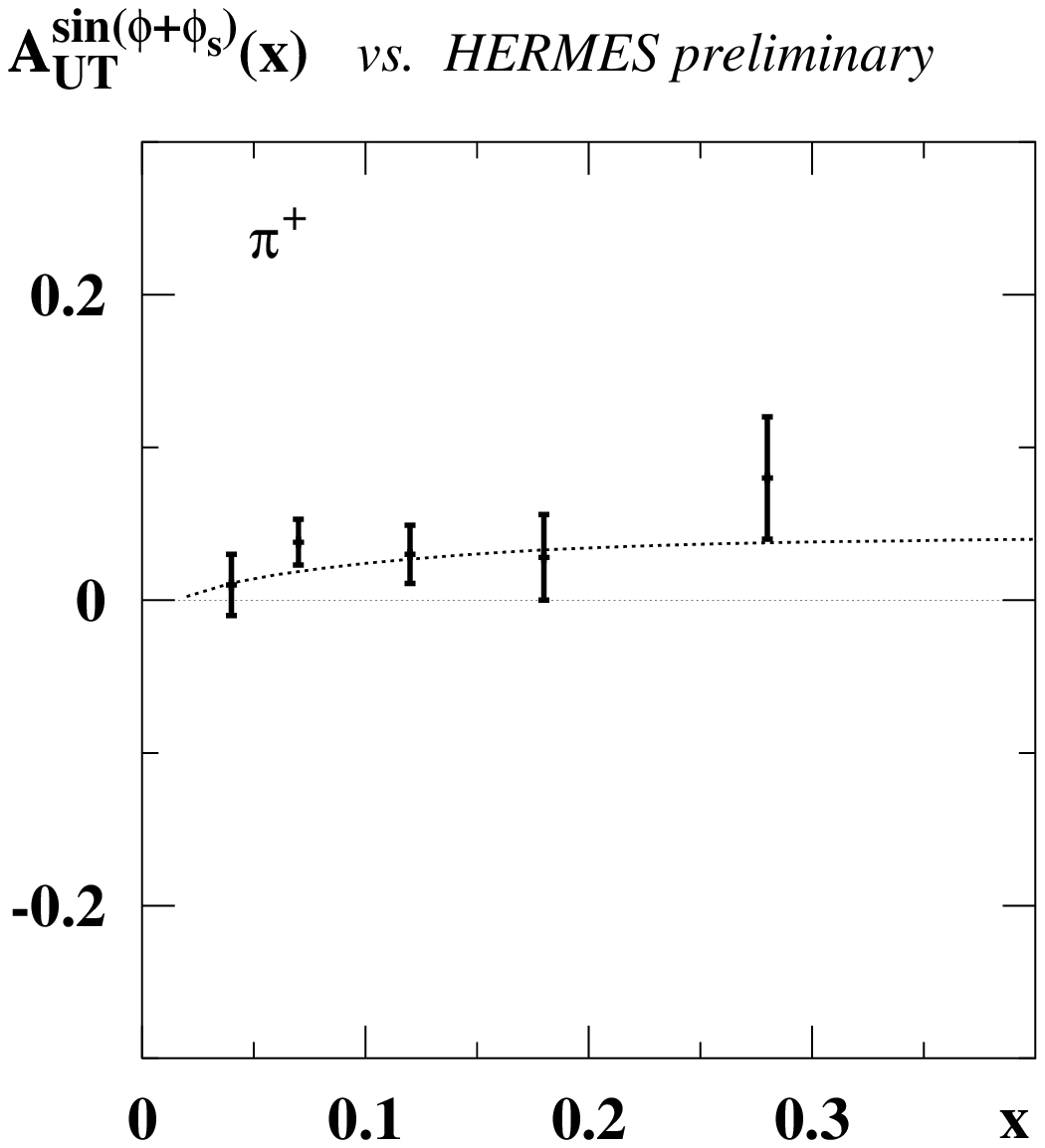}
\includegraphics[width=.30\textwidth]
{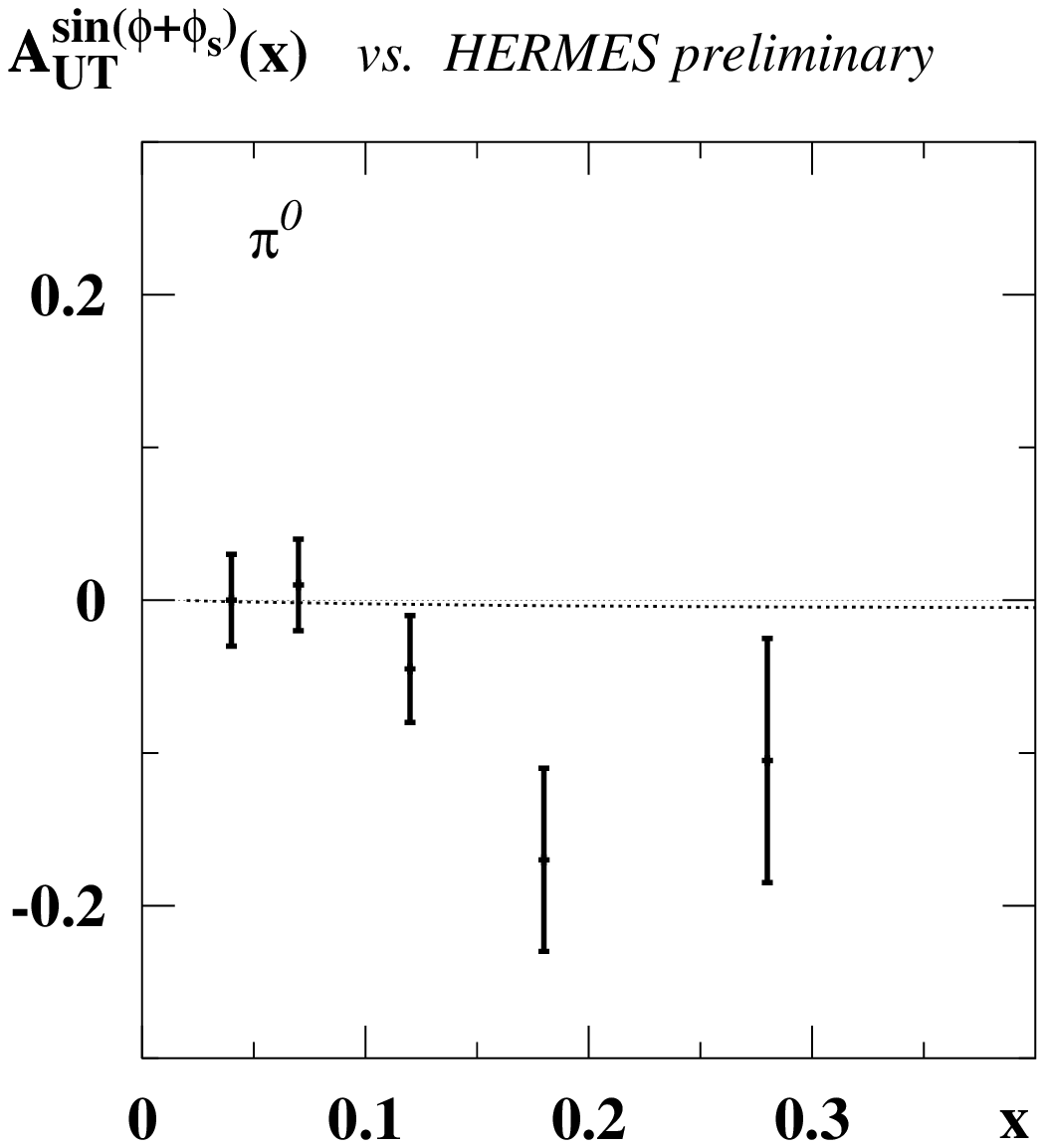}
\includegraphics[width=.30\textwidth]
{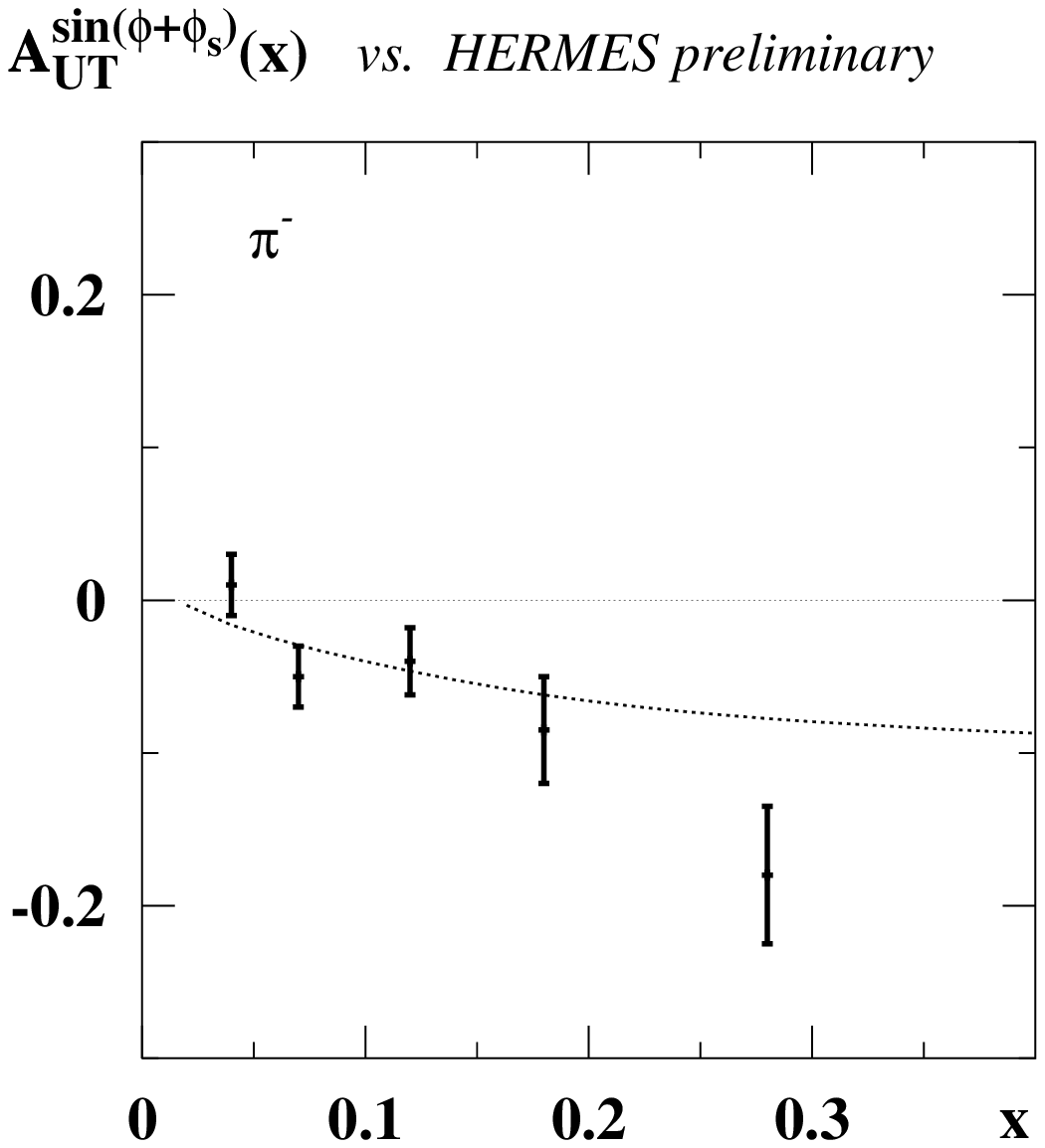}
\end{center}
\vspace{-0.8cm}
\caption[*]{
The Collins effect transverse target
SSA $A_{0T}^{\sin(\phi+\phi_s)}$ in the production of $\pi^+$,
$\pi^0$ and $\pi^-$ from a proton target. {\sl Preliminary} data
are from \cite{AUT-HERMES}, theoretical curves consider
unfavoured fragmentation effects assuming $\la H_1^{\perp\rm
unf}\ra = -1.2 \, \la H_1^{\perp\rm fav}\ra < 0$, and $h_1^a(x)$
is taken from \cite{Schweitzer:2001sr}. The description improves
for charged pions, but it fails for the neutral pion. ${}\;\;{}$
\label{AUT-col-HERMES-unfH1-1.2}}
\begin{center}
\includegraphics[width=.30\textwidth]
{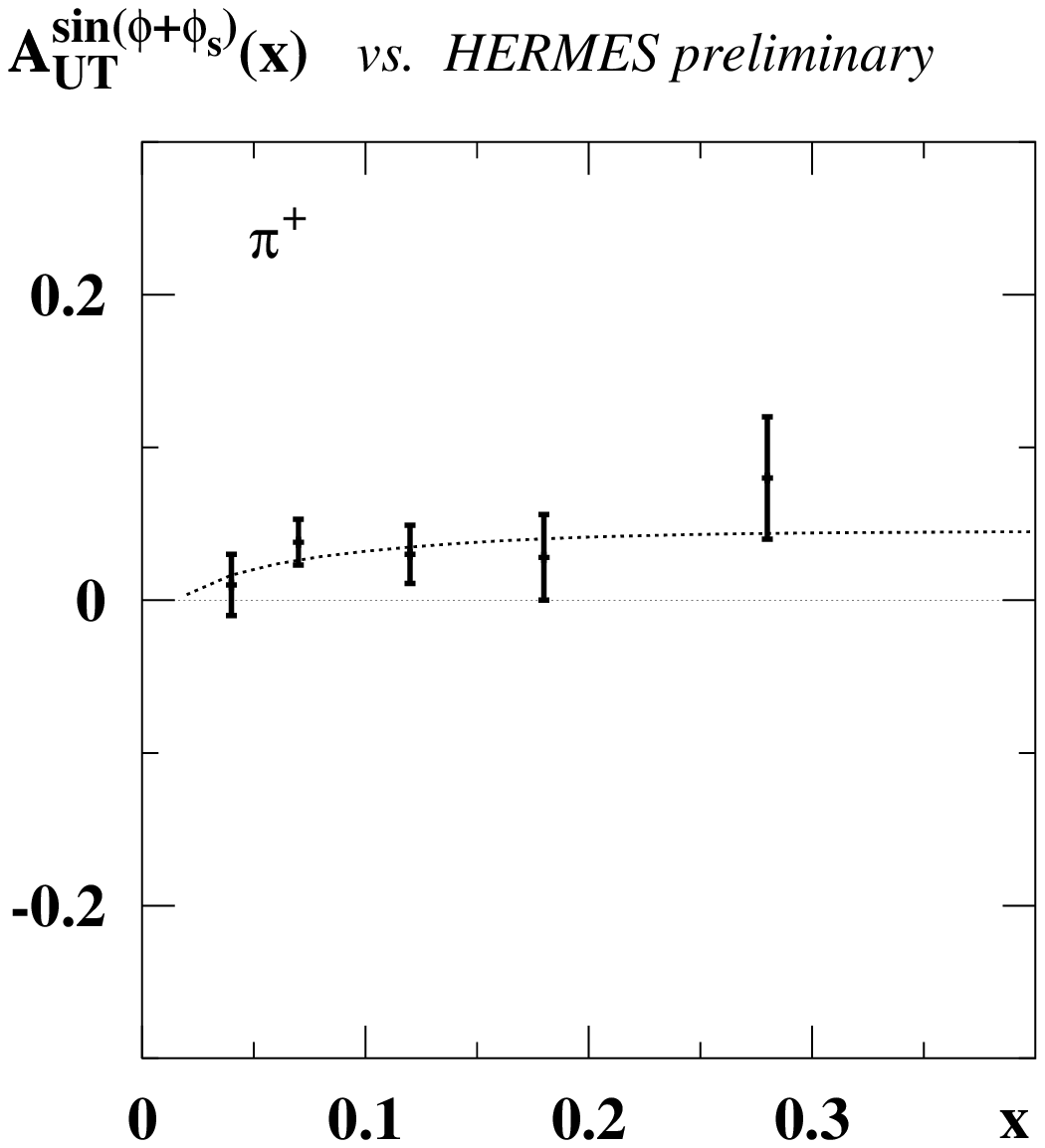}
\includegraphics[width=.30\textwidth]
{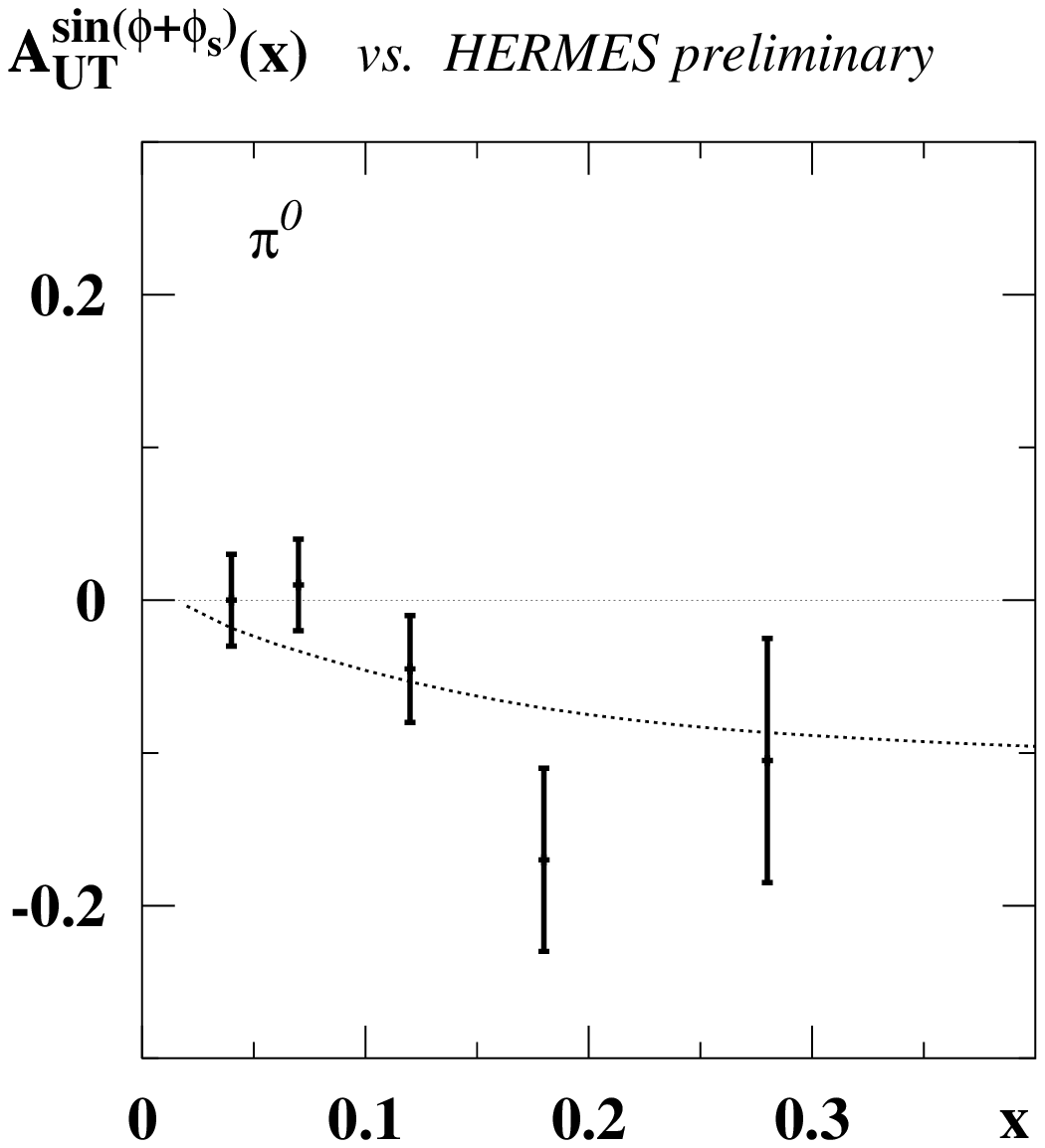}
\includegraphics[width=.30\textwidth]
{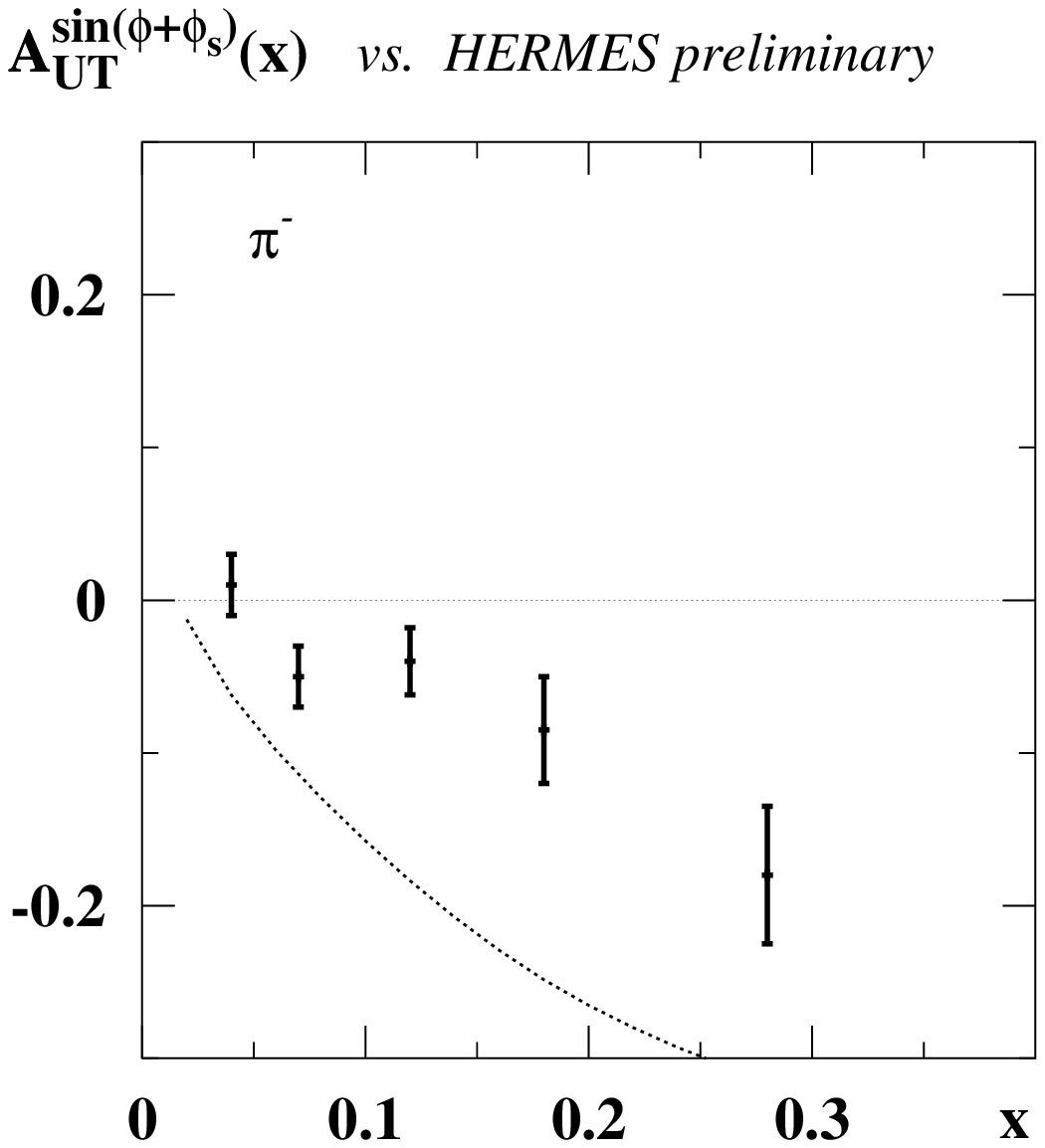}
\end{center}
\vspace{-0.8cm}
\caption[*]{
The same as Fig.~\ref{AUT-col-HERMES-unfH1-1.2} but assuming
$\la H_1^{\perp\rm unf}\ra = -5 \la H_1^{\perp\rm fav}\ra < 0$.
Now the description is reasonable for $\pi^+$ and $\pi^0$, however,
the $\pi^-$ asymmetry cannot be described in this way.
\label{AUT-col-HERMES-unfH1-5}}
\vspace{-0.5cm}
\end{figure}

Let us illustrate the problem from another point of view by
considering tentatively the effect of unfavoured fragmentation.
Let us assume $\la H_1^{\perp\rm fav}\ra = 0.0215$ which would
correspond to Eq.~(\ref{apower}) with unfavoured fragmentation
neglected\footnote{ Of course, if unfavoured fragmentation is
relevant at HERMES, then it was also relevant at DELPHI. However,
all we are interested in here, is to fix numbers and get a rough
insight.}, and $\la H_1^{\perp\rm unf}\ra\approx -1.2\;\la
H_1^{\perp\rm fav}\ra$ which is compatible with the observation
in Eq.~(\ref{AUT-unf-pi0}). With $\la D_1^{\rm fav}\ra$ and $\la
D_1^{\rm unf}\ra$ from Ref.~\cite{Kretzer:2001pz} we obtain the
results plotted in Fig.~\ref{AUT-col-HERMES-unfH1-1.2}. Clearly,
the description of charged pions improves and looks reasonable.
However, it is not possible to describe the $\pi^0$ data in this
way. Although the sign of the $\pi^0$ asymmetry is now correct,
the magnitude of the effect is strongly underestimated. If one
assumed $\la H_1^{\perp\rm unf}\ra\approx -5\;\la H_1^{\perp\rm
fav}\ra$ (How to explain in any string model!?), than $\pi^+$ and
$\pi^0$ asymmetries would be reasonably described. However, then
the $\pi^-$ asymmetry would be overestimated by factor of two,
see Fig.~\ref{AUT-col-HERMES-unfH1-5}. And, of course, we totally
loose the description of $A_{0L}^{\sin\phi}$ for $\pi^0$ and
$\pi^-$.

Although our estimates are rough, we nevertheless learn an
important lesson. It seems not possible to describe the
preliminary HERMES data by considering unfavoured fragmentation,
at least not on the basis of $h_1^a(x)$ from the $\chi$QSM
\cite{Schweitzer:2001sr}. Instead, one should consider in
addition a different transversity distribution, where $h_1^d$
would dominate over $h_1^u$. This would be an unexpected scenario
from the point of view of actually any model of $h_1^a$, and
unusual from the point of view of the experience with the other
twist-2 nucleon distribution functions $f_1^a$ and $g_1^a$.

The present situation seems paradoxical. We have a reasonable
understanding of $A_{0L}$ asymmetries, but know that it possibly
is based on an incomplete theoretical description of the process
-- with the Sivers effect and other contributions omitted. We
probably have a more reliable description of the Collins $A_{0T}$
asymmetry, but cannot understand the {\sl preliminary} data.

However, one should keep in mind the preliminary stage of the
data \cite{AUT-HERMES}, which does not allow yet to draw more
definite conclusions. Further data from HERMES as well as
COMPASS, CLAS, HALL-A and HALL-B experiments will contribute
considerably to resolve the present puzzles and pave the way
towards a qualitative understanding of the numerous new
distribution and fragmentation functions.

\section{Transversity from polarized $\bm p\bar p$ Drell-Yan
process}

The Drell-Yan process remains up to now the theoretically
cleanest and safest way to access $h_1^a(x)$. The first attempt
to study $h_1^a(x)$ by means of the Drell-Yan process is planned
at RHIC. Dedicated estimates, however, indicate that at RHIC the
access of $h_1^a(x)$ by means of the Drell-Yan process is very
difficult. The main reason is that the observable double spin
asymmetry $A_{TT}$ is proportional to a product of transversity
quark and antiquark PDF. The latter are small, even if they were
as large as to saturate the Soffer upper limit \cite{Soffer:1995ww}.

This problem can be circumvented by using an antiproton beam.
Then $A_{TT}$ is proportional to a product of transversity quark
PDF from the proton and transversity antiquark PDF from the
antiproton (which are equal due to charge conjugation). Thus in
this case one can expect sizeable counting rates. The challenging
program how to polarize an antiproton beam has been recently
suggested in the PAX experiment at GSI \cite{PAX}. The
technically realizable polarization of the antiproton beam of
$(5-10)\%$ \cite{Rathmann:2004pm} and the large counting rates
make the program rather promising. Here we shortly describe our
quantitative estimates for the Drell-Yan double spin asymmetry
$A_{TT}$ in the kinematics of the PAX experiment at LO QCD. (For
more details and references see \cite{Efremov:2004qs}.)

The process $p\bar p\to \mu^+\mu^-X$ can be characterized by the
invariants: Mandelstam $s=(p_1+p_2)^2$ and dilepton invariant
mass $Q^2=(k_1+k_2)^2$, where $p_{1/2}$ and $k_{1/2}$ are the
momenta of respectively the incoming proton-antiproton pair and
the outgoing lepton pair, and the rapidity $y=\frac12\,{\rm
ln}\frac{p_1(k_1+k_2)}{p_2(k_1+k_2)}$. The double spin asymmetry
in Drell-Yan process is given by
\be\label{Eq:ATT-0}
\frac{N^{\uparrow\uparrow}-N^{\uparrow\downarrow}}
{N^{\uparrow\uparrow}+N^{\uparrow\downarrow}}
= D_P \; \frac{\sin^2\theta}{1+\cos^2\theta}\;\cos 2\phi\;
A_{TT}(y,Q^2) \;,
\ee
where $\theta$ is the emission angle of one lepton in the
dilepton rest frame and $\phi$ its azimuth angle around the
collision axis counted from the polarization plane of the hadron
whose spin is not flipped in Eq.~(\ref{Eq:ATT-0}). The factor
$D_P$ takes into account polarization effects. At LO QCD $A_{TT}$
is given by
\be
\label{Eq:ATT-1}
A_{TT}(y,Q^2)=\frac{\sum_a e_a^2h_1^a(x_1,Q^2) h_1^a(x_2,Q^2)}
{\sum_b e_b^2 f_1^b(x_1,Q^2)f_1^b(x_2,Q^2)}\;,
\ee
where the parton momenta $x_{1/2}$ in Eq.~(\ref{Eq:ATT-1}) are
$x_{1/2} = \sqrt{\frac{Q^2}{s}}\,e^{\pm y}$. In
Eq.~(\ref{Eq:ATT-1}) use was made of the charge conjugation
invariance. For the transversity distribution we shall use
predictions from the chiral quark soliton
model \cite{Schweitzer:2001sr}.

\begin{figure}[ht!]
\begin{center}
\vspace{-7mm}
\epsfxsize=5cm\epsfbox{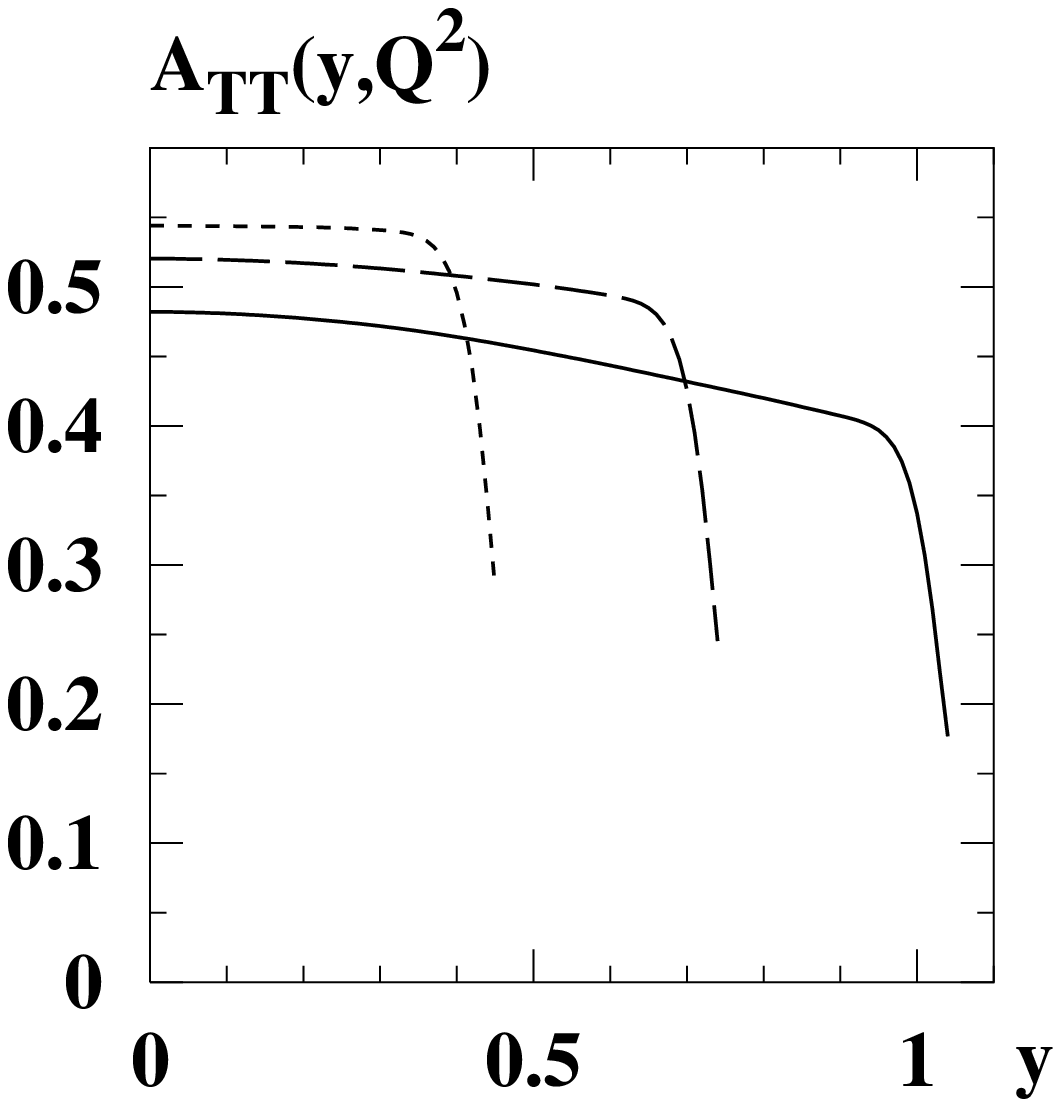} {\bf a)}
\epsfxsize=5cm\epsfbox{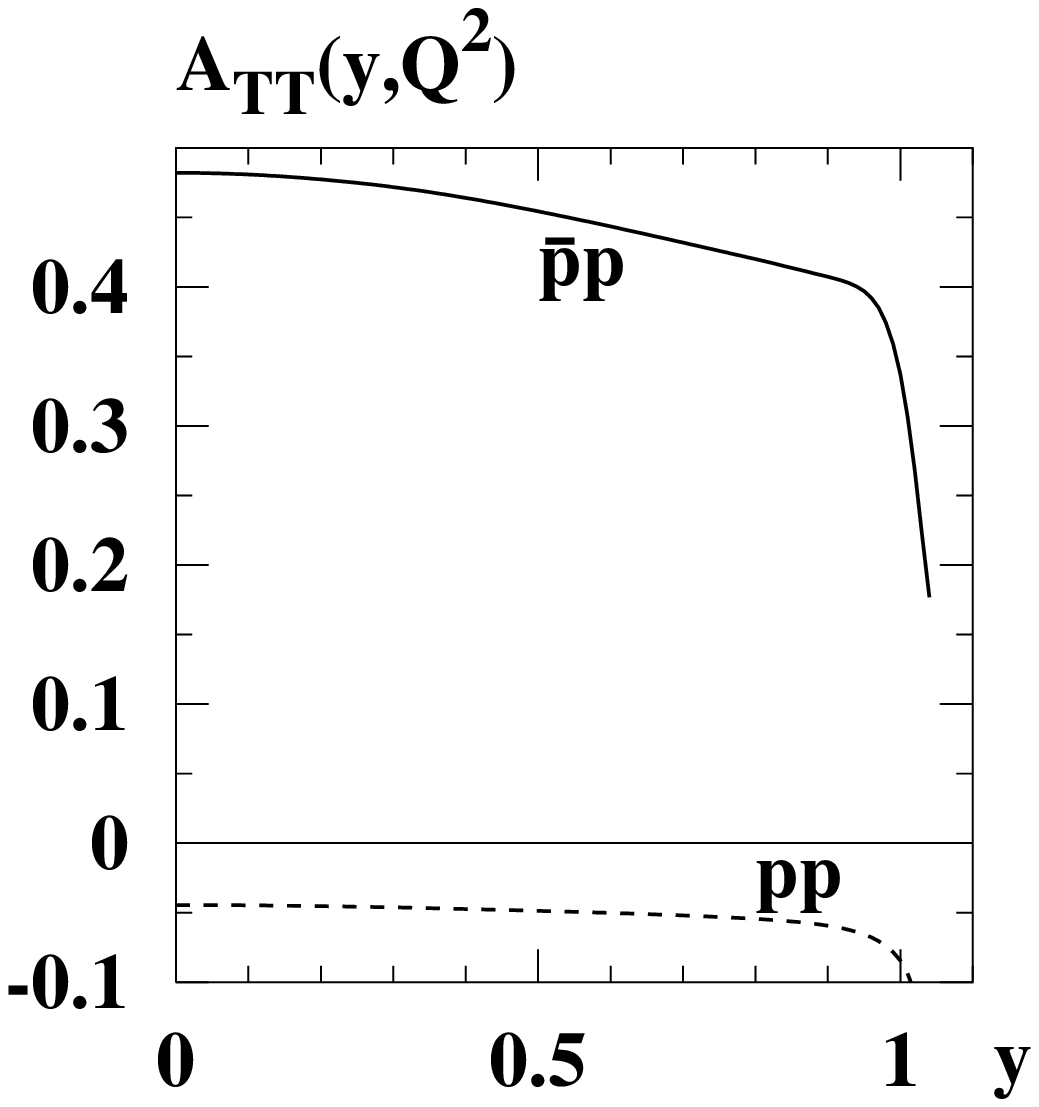} {\bf b)}
\end{center}
\caption{
{\bf a)}
The asymmetry $A_{TT}(y,M^2)$, {\sl cf.} Eq.~(\ref{Eq:ATT-1}),
as function of the rapidity $y$ for $Q^2=5\,{\rm GeV}^2$ (solid)
and $9\,{\rm GeV}^2$ (dashed) and $16\,{\rm GeV}^2$ (dotted line)
for $s=45\,{\rm GeV}^2$.
{\bf b)}
Comparison of $A_{TT}(y,M^2)$ from proton-antiproton
(solid) and proton-proton (dotted line) collisions at PAX for
$Q^2=5\,{\rm GeV}^2$ and $s=45\,{\rm GeV}^2$.
}
\label{figpax}
\end{figure}

In the PAX experiment an antiproton beam with energies in the
range $(15-25)\,{\rm GeV}$ could be available, which yields
$s=(30-50)\,{\rm GeV}^2$ for a fixed proton target. The region
$1.5\,{\rm GeV} < Q < 3\,{\rm GeV}$, i.e. below the $J/\Psi$
threshold but well above $\Phi(1020)$-decays (and with
sufficiently large $Q^2$) would allow to explore the region
$x>0.2$. However, in principle one can also address the resonance
region itself and benefit from large counting
rates \cite{Anselmino:2004ki} since the unknown $q\bar q J/\Psi$
and $J/\Psi\mu^+\mu^-$-couplings cancel in the ratio in
Eq.~(\ref{Eq:ATT-0}) as argued in Ref.\cite{Anselmino:2004ki}.
Keeping this in mind we shall present below estimates for
$s=45\,{\rm GeV}^2$, and $Q^2=5\,{\rm GeV}^2$, $9\,{\rm GeV}^2$
and $16\,{\rm GeV}^2$ (see Fig.\ref{figpax}) .

The exploitable rapidity range shrinks with increasing dilepton
mass $Q^2$. Since $s=x_1x_2Q^2$, for $s=45\,{\rm GeV}^2$ and
$Q^2=5\,{\rm GeV}^2$ ($16\,{\rm GeV}^2$) one probes parton
momenta $x>0.3$ ($x>0.5$). The asymmetry $A_{TT}$ grows with
increasing $Q^2$ where larger parton momenta $x$ are involved,
since $h_1^u(x)$ is larger with respect to $f_1^u(x)$ in the
large $x$-region (see Fig.\ref{fig-h1}).

The advantage of using antiprotons is evident from
Fig.\ref{figpax}b. The corresponding asymmetry from proton-proton
collisions is an order of magnitude smaller (this observation
holds also in the kinematics of RHIC \cite{Schweitzer:2001sr}).
At first glance this advantage seems to be compensated by the
polarization factor in Eq.~(\ref{Eq:ATT-0}). For the antiproton
beam polarization of $(5-10)\%$ and the proton target
polarization of $90\%$, i.e.\ at PAX $D_P\approx 0.05$. However,
thanks to the use of antiprotons the counting rates are more
sizeable. A precise measurement of $A_{TT}$ in the region
$Q>4\,{\rm GeV}$ is very difficult, however, in the dilepton mass
region below the $J/\Psi$ threshold \cite{PAX} and in the
resonance region \cite{Anselmino:2004ki} $A_{TT}$ could be
measured with sufficient accuracy in the PAX experiment.

A precise measurement would allow to discriminate between
different models for $h_1^a(x)$, e.g., the popular guess
$h_1^a(x)\approx g_1^a(x)$ where one would expect
\cite{Anselmino:2004ki} the $A_{TT}$ of about $30\%$ to be
contrasted with the chiral quark soliton model estimate of about
$50\%$.

At next-to-leading order in QCD one can expect corrections to
this result which reduce somehow the asymmetry \cite{NLO}.

\section{Conclusion}

Processes with polarized particles have always been among the
most difficult and complicated themes both for experimentalists
and theorists.

First, working with polarized targets, experimentalists have to
"battle with" thermal chaos which tends to break the polarized
order. For this one needs liquid helium temperatures. More
difficulties, like depolarizing resonances, are encountered in
accelerating polarized particles and in controlling a polarized
beam. Second, spin effects are very perfidious: as a rule, they
are strongest in kinematical regions where the process itself is
the least probable.

As for the theory, one can hardly recall a case when its first
prediction was correct! As a rule, it was wrong and forced
theorists to think more fundamentally to repair the theory. This
resulted in a deeper understanding of particle interaction
mechanics. Nevertheless many puzzles such as "Why are hyperons
produced so strongly polarized?" or "What is the structure of the
nucleon spin?" stay yet unsolved during decades. Will single spin
asymmetries become one more such problem? The future will show.

 \end{document}